%% file: main.tex
\documentclass[acmsmall,anonymous=false,natbib=true,screen=true, review=false]{acmart}
\usepackage[utf8]{inputenc}

\input{packages}
\input{definitions}
\input{meta}
\begin{document}





\title[Do Generative Recommenders Deepen the Information Cocoon?
]{Do Generative Recommenders Deepen the Information Cocoon? A Closed-Loop Simulation with LLM-powered User Simulators}


\input{sections/abstract.tex}


\maketitle
\acresetall

\input{sections/introduction.tex}

\input{sections/method.tex}
\input{sections/metrics.tex}
\input{sections/experiments.tex}

\input{sections/discussion.tex}

\input{sections/related_work.tex}
\input{sections/conclusion.tex}
\clearpage
\bibliographystyle{ACM-Reference-Format}
\balance
\bibliography{main}
\clearpage
\nobalance
\onecolumn

\input{sections/appendix.tex}
\clearpage
\end{document}

%% file: packages.tex
\usepackage{booktabs} 
\usepackage{amsmath}
\usepackage{graphics}
\usepackage{epsfig}
\usepackage{graphicx}
\usepackage{xcolor}
\usepackage{balance}
\usepackage{multirow}
\usepackage{mathrsfs}
\usepackage{acronym}
\usepackage{placeins}
\usepackage{color}
\usepackage[inline]{enumitem}
\usepackage{fancyhdr}
\usepackage{amsfonts}
\usepackage{svg}
\usepackage{threeparttable}
\usepackage{subfig}
\usepackage{float}
\usepackage{bbm}
\usepackage{array}
\usepackage{amssymb}
\usepackage{amsthm}
\usepackage[title]{appendix}
\usepackage{amsmath}  
\usepackage{amssymb}  
\usepackage{mathrsfs} 

\usepackage{algorithm}
\usepackage{algorithmicx}
\usepackage{algpseudocode}  
\usepackage[most]{tcolorbox}
\usepackage{listings}
\usepackage{cuted}
\usepackage{todonotes}
\newcounter{promptcnt}
\usepackage{subfig}
\newtcblisting{prompt}[1][Prompt Template]{%
    enhanced,
    breakable,
    colback=gray!5,
    colframe=gray!40,
    listing only,
    listing options={
        basicstyle=\ttfamily\footnotesize,
        columns=flexible,
        breaklines=true,
        keepspaces=true,
        breakindent=0em,
        escapechar=\%,
        upquote=true
    },
    title={#1},
    attach boxed title to top left={xshift=0.5cm, yshift=-\tcboxedtitleheight/2},
    boxed title style={colback=gray!40, colframe=gray!40, sharp corners},
    fonttitle=\bfseries\small,
    before upper={\refstepcounter{promptcnt}},
    title after break={#1~(cont'd)},
    lines before break=4,
}

\tcbuselibrary{skins, breakable}
\usepackage[dvipsnames]{xcolor}

%% file: definitions.tex
\AtBeginDocument{%
  \providecommand\BibTeX{{%
    \normalfont B\kern-0.5em{\scshape i\kern-0.25em b}\kern-0.8em\TeX}}}

\newtheorem{definition}{Definition}[section]
\usepackage[most]{tcolorbox}
\definecolor{insightblue}{HTML}{E5F0F9} 

\newtcolorbox{insightbox}{
  colback=insightblue,       
  colframe=insightblue,      
  arc=2.5mm,                 
  boxrule=0pt,               
  enhanced,
  breakable,
  top=3mm, bottom=3mm,       
  left=3mm, right=3mm,
  boxsep=0pt
}

%% file: meta.tex
\setcopyright{acmcopyright}
\acmYear{2026} \acmVolume{1} \acmNumber{1} \acmArticle{1} \acmMonth{1}
\acmPrice{15.00}
\acmJournal{TOIS}

\author{Jiyuan Yang}
\email{jiyuan.yang@mail.sdu.edu.cn}
\affiliation{%
  \institution{Shandong University}
  \city{Qingdao}
  \country{China}
}

\author{Gengxin Sun}
\email{gxin.sun@mail.sdu.edu.cn}
\affiliation{%
  \institution{Shandong University}
  \city{Qingdao}
  \country{China}
}

\author{Mengqi Zhang}
\email{mengqi.zhang@sdu.edu.cn}
\affiliation{
  \institution{Shandong University, Shandong Provincial Key Laboratory of Computing-Network Integration, Qingdao Key Laboratory of Trustworthy Artificial Intelligence}
  \city{Qingdao}
  \country{China}
}
\author{Lingjie Wang}
\email{lingjie.wang@mail.sdu.edu.cn}
\affiliation{%
  \institution{Shandong University}
  \city{Qingdao}
  \country{China}
}

\author{Yuanzi Li}
\email{liyuanzi0313@outlook.com}
\affiliation{%
  \institution{Renmin University of China}
  \city{Beijing}
  \country{China}
}

\author{Hongxi Cui}
\email{cuihongxi@mail.sdu.edu.cn}
\affiliation{%
  \institution{Shandong University}
  \city{Qingdao}
  \country{China}
}

\author{Xin Xin}
\email{xinxin@sdu.edu.cn}
\affiliation{%
  \institution{Shandong University}
  \city{Qingdao}
  \country{China}
}

\author{Pengjie Ren}
\authornote{Corresponding author.}
\email{renpengjie@sdu.edu.cn}
\affiliation{%
  \institution{Shandong University, Shandong Provincial Key Laboratory of Computing-Network Integration, Qingdao Key Laboratory of Trustworthy Artificial Intelligence}
  \city{Qingdao}
  \country{China}
}

\renewcommand{\shortauthors}{Jiyuan Yang and Gengxin Sun et al.}

%% file: sections/abstract.tex
\begin{abstract}
Recommender systems alleviate information overload by filtering large-scale content, yet repeated feedback between recommendations and user interactions can reinforce users' existing preferences and progressively narrow the range of content users receive, forming information cocoons.
While information cocoons have been extensively studied in traditional ID-based sequential recommendation, their effects in generative recommendation remain largely unclear.
Unlike conventional sequential recommenders that score candidate items represented by atomic IDs, generative recommenders represent items as discrete code sequences and generate recommendations autoregressively.
This shift from atomic-ID scoring to code-sequence generation makes it unclear whether generative recommenders mitigate, inherit, or deepen information cocoon effects.

To answer this question, we present \textsc{RecLoop}, a large-scale closed-loop simulation framework that enables long-term analysis of information cocoon formation in generative recommendation.
\textsc{RecLoop} runs two generative recommenders and two traditional sequential baselines with LLM-driven user agents on two Amazon product datasets, involving 5K agents on Office Products and 20K agents on Toys \& Games, over 15 feedback cycles with periodic model retraining.
To reflect the sequential recommendation setting, user agents are initialized from historical behavior sequences and update their preference states as new interactions occur.
We evaluate exposure-level cocoon formation with standard metrics that capture individual exposure narrowing, cross-user homogenization, and system-wide exposure concentration.
Beyond observable exposure outcomes, we introduce \emph{Code-Space Structural Cocoon}, a model-level metric measures whether the generated code space of generative recommenders becomes increasingly concentrated over feedback cycles.

Our analysis shows that generative recommenders are generally less prone to exposure-level cocoon formation than traditional sequential baselines, preserving broader user exposure and slowing cross-user homogenization.
However, generative recommenders are not immune to cocoon effects: feedback loops can still induce concentration within the generated code space.
We further find that cocoon severity is strongly shaped by tokenization strategy and model scale.
Collaborative-signal tokenization leads to stronger cocoon effects than semantic tokenization, while larger models preserve broader code-space diversity and better retain access to niche content.
These results suggest that information cocoons in generative recommendation are governed not only by retrieval accuracy, but also by how items are tokenized and how much generative capacity the model has.
Our code is available at \url{https://github.com/Dregen-Yor/RecLoop}.
\end{abstract}

%% file: sections/introduction.tex
\section{INTRODUCTION}
\begin{figure}[ht]
    \centering
    \includegraphics[width=1.0\linewidth]{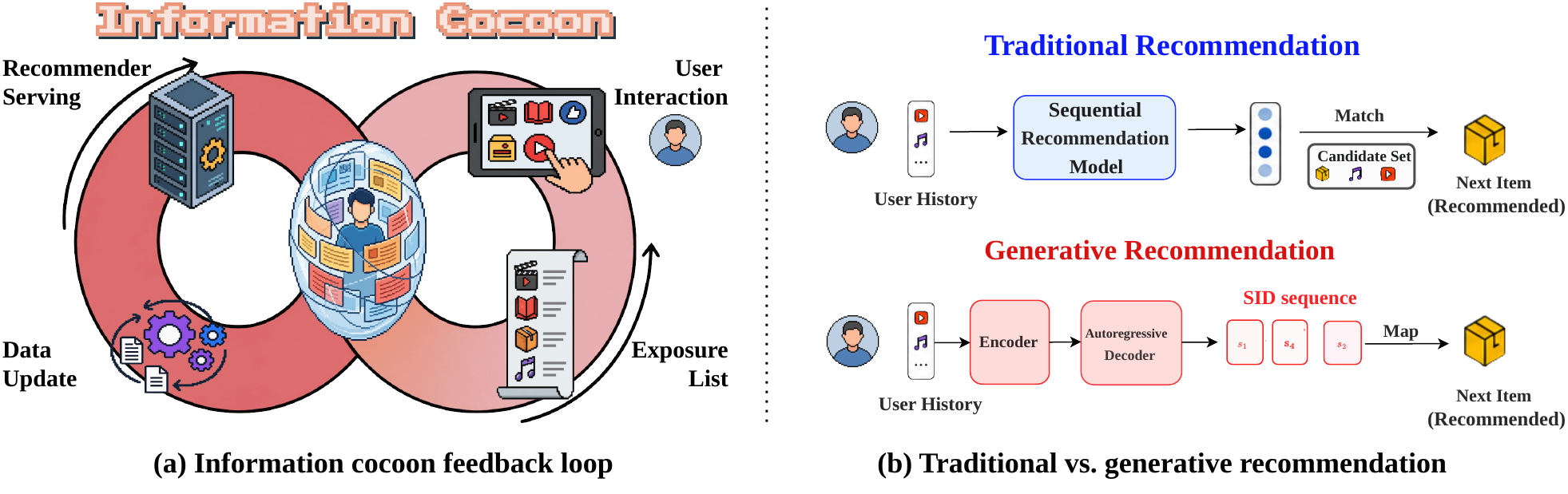}
    \caption{Overview of information cocoon formation and recommendation paradigms. (a) A closed feedback loop in which recommendation, user interaction, and data update iteratively reinforce exposure patterns. (b) Traditional sequential recommenders match predicted next-item representations with atomic item IDs, whereas generative recommenders autoregressively generate SID sequences and map them back to items.}
    \label{fig:framework}
\end{figure}

Recommendation is not only about finding what users may like, but also about deciding what information they are likely to encounter. When a user watches several anime clips, the system may infer a stronger preference for anime, recommend more anime-related content, and collect even more anime-related feedback shown in Figure~\ref{fig:framework} (a). After several rounds, the feed may become highly relevant but surprisingly narrow, with topics such as travel or science gradually disappearing. This phenomenon, often known as an \emph{information cocoon}, reflects a self-reinforcing loop in which recommender systems optimize for inferred preferences while progressively reducing the diversity of user exposure~\cite{castells2021novelty, pariser2011filter, peng2021breaking, li2022exploratory, liang2023influence, zhang2025exploratory, he2024information, zhaoziqi2026improving, yang2024debiasing}.
Information cocoons have been widely studied in traditional recommender systems, where repeated feedback loops can reinforce user interests, amplify popularity bias, and reduce exposure diversity~\cite{lin2025recommendation, 10.1145/3306618.3314288}. However, the recent rise of generative recommendation makes this problem worth revisiting. Generative recommenders have rapidly evolved from an emerging research direction to systems with large-scale industrial deployment~\cite{tiger, zhou2025onerec, deng2025onerec, letter}. Despite this progress, their long-term information cocoon effects remain largely unexplored. This leaves an important question unanswered: 
\textit{do generative recommenders inherit, mitigate, or even amplify the information cocoon effects observed in traditional recommendation?}

\noindent\textbf{Information cocoons in generative recommendation.}
The above question is non-trivial because generative recommendation changes how items are represented and produced. As shown in Figure~\ref{fig:framework} (b), generative recommenders tokenize items into sequences of discrete codes and produce recommendations by autoregressively generating such code sequences~\cite{tiger, zhou2025onerec, deng2025onerec, letter} instead of directly scoring a fixed candidate set. These item tokens can be constructed by different quantization or clustering methods, but they commonly introduce a structured discrete representation space. This changes how exposure may be concentrated or diversified. Because multiple items can share tokens or code prefixes, a shift toward one frequently generated code may affect an entire region of the item space rather than a single item. Meanwhile, multi-layer token generation makes later codes depend on earlier, coarser-level decisions, so narrowing at an early layer may propagate to subsequent layers.

These structural properties make the information cocoon effect in generative recommendation difficult to predict in advance. A shared token space may concentrate exposure if the model repeatedly generates a small set of popular or semantically dominant codes, but it may also spread exposure across multiple items that share related codes. Similarly, hierarchical generation may preserve diversity at coarse levels, but it may also amplify narrowing when early code choices constrain later decoding steps. Therefore, whether generative recommendation mitigates or exacerbates information cocoons is an empirical question. 
To answer this question, we develop a closed-loop simulation framework, quantify information cocoons from both exposure-level and generative-code perspectives, and conduct experiments to compare how different recommendation models shape cocoon dynamics over long-term feedback cycles.
\noindent\textbf{Simulation framework.}
We first build a user simulator that models users' long-term responses to recommendation exposure, and then introduce \textsc{RecLoop}, a closed-loop simulation framework for analyzing information cocoon effects.
Information cocoons are inherently long-term phenomena that emerge through repeated feedback cycles: the system's recommendations shape user exposure, user interactions update the training data, and the updated data further affects subsequent recommendations.
Such feedback effects cannot be fully captured by standard offline evaluation protocols, where user preferences and exposure distributions are typically treated as fixed.
This limitation is especially important in sequential recommendation, where recent interactions provide strong signals about users' current interests.
Therefore, a faithful simulator should not only preserve users' historical preferences, but also promptly update their preference states as new interactions occur.

Following recent work on LLM-based user simulation~\cite{zhang2024generative, chen2025recusersim, zhang2025llm}, \textsc{RecLoop} uses LLM-powered user agents to simulate repeated recommendation--interaction--retraining cycles.
At each cycle, the recommender serves an exposure list, the user agent selects items according to its current preference state, and the selected items are recorded as new interactions, appended to the user's behavior sequence, and added back to the training data for periodic model retraining.
To support sequential preference evolution, each agent maintains both long-term historical interests and short-term interaction context, allowing it to capture emerging interests while remaining grounded in the user's original behavior.

\noindent\textbf{Quantifying cocoon formation.}
Based on the simulated feedback cycles, we quantify information cocoon effects from both exposure-level and model-level perspectives.
For exposure-level evaluation, we adopt standard metrics to characterize three common symptoms of cocoon formation: whether each user's exposure range becomes narrower, whether different users receive increasingly similar recommendations, and whether system-wide exposure becomes concentrated on a smaller portion of the item space~\cite{diversity,stamenkovic2021choosing,anwar2024homogenization,fleder2009blockbuster,areeb2023filter,sukiennik2024deep,bellogin2013comparative}.
These metrics provide a basis for comparing traditional and generative recommenders, but they only capture cocoon effects from observed exposure outcomes.

To further characterize information cocoons from the perspective of generative recommendation itself, we introduce \emph{Code-Space Structural Cocoon}, a generative-specific model-level metric defined over generated code representations.
Since generative recommenders represent items as multi-layer code sequences, cocoon formation may appear not only as repeated exposure to similar items or categories, but also as reduced diversity in the discrete code space used by the model.
\emph{Code-Space Structural Cocoon} measures how concentrated the generated codes become as feedback cycles proceed, capturing whether the model's generative space itself becomes narrower over time.
Unlike standard exposure-level metrics, this metric describes information cocoon effects within the representation and generation space of generative recommenders.
By integrating exposure-based metrics with our proposed code-space metric, we are able to examine both the visible effects of information cocoons and how they are reflected at the model level within generative recommendation.

\noindent\textbf{Experiments and Key findings.}
We conduct a large-scale closed-loop simulation in which two generative recommenders and two traditional sequential baselines interact with LLM-driven user agents.
The simulation runs on two Amazon product datasets, involving 5K agents on Office Products and 20K agents on Toys \& Games, over 15 feedback cycles with periodic model retraining.
Our analysis leads to four main findings:

\begin{itemize}[leftmargin=*,nosep]
\item \textbf{Generative recommenders are less prone to exposure-level cocoons.} Under the same closed-loop feedback protocol, generative recommenders generally preserve broader individual exposure diversity and exhibit slower inter-user homogenization than traditional sequential baselines.
\item \textbf{Cocoon effects also appear in the generative code space.} Beyond observable exposure outcomes, our code-space analysis shows that feedback loops can reduce diversity within generated code representations. This suggests that information cocoons in generative recommendation should be examined not only at the item or category level, but also at the model-representation level.
\item \textbf{Tokenization affects cocoon severity.} Different item tokenization strategies lead to different cocoon dynamics. In our experiments, collaborative ID tokenization tends to produce stronger cocoon effects than semantic ID tokenization, suggesting that collaborative signals may carry popularity bias into the discrete code space.
\item \textbf{Model scale helps preserve diversity.} Larger generative recommenders maintain a broader active code space during closed-loop simulation and are less likely to lose access to niche content, indicating that model capacity can buffer cocoon formation.
\end{itemize}

\noindent Overall, this work provides a systematic analysis of information cocoons in generative recommendation. By combining closed-loop simulation, exposure-level evaluation, and code-space analysis, we show that generative recommenders exhibit cocoon dynamics that differ from traditional sequential models and deserve dedicated attention beyond standard retrieval accuracy.

%% file: sections/method.tex
\section{SIMULATION FRAMEWORK}
\label{sec:framework}

\begin{figure*}[ht]
    \centering
\includegraphics[width=\linewidth]{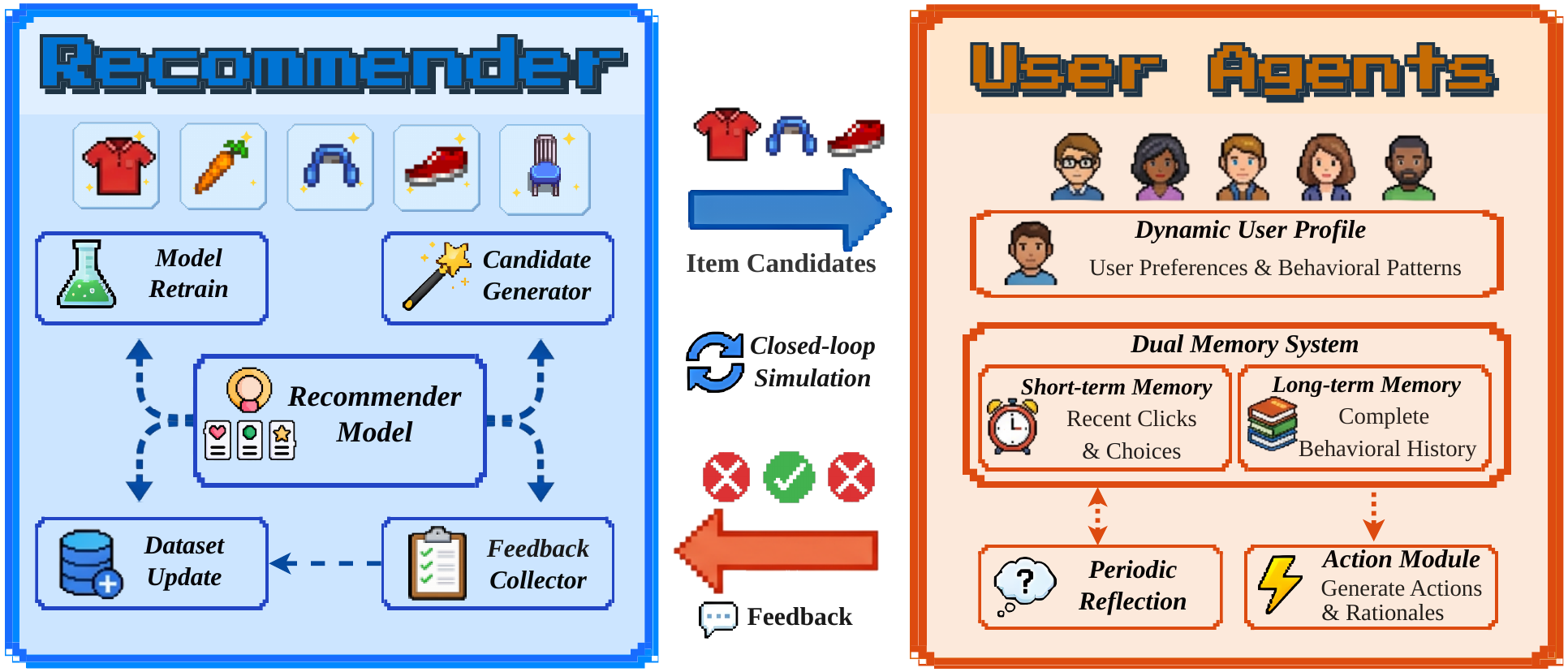}
    \caption{Overview of the \textsc{RecLoop} closed-loop simulation framework.
    The framework couples a recommender module for candidate generation with an LLM-powered user agent module for simulating user behavior, establishing a closed feedback loop between recommendation exposure and simulated user feedback.}
    \label{fig:recloop}
\end{figure*}

Studying information cocoon formation requires a closed-loop environment in which user feedback is repeatedly absorbed by the recommender and further shapes subsequent exposure.
Such feedback loops are difficult to capture through static offline evaluation, where recommenders are evaluated on fixed historical logs and the feedback induced by model-generated exposure is not returned to the system.
The most direct way to study such dynamics is to deploy different recommenders to real users and observe how their exposure and feedback evolve over multiple cycles.
However, such a multi-cycle user study is difficult to conduct, reproduce, and control at scale, because the experimental algorithms themselves continuously reshape the information environment from which subsequent user feedback is collected.
Motivated by recent studies~\cite{zhang2024generative, zhang2024agentcf} showing that LLM-powered agents can serve as effective user simulators, we construct \textsc{RecLoop}, a closed-loop simulation framework for tracing how recommendation exposure, simulated user feedback, and model updates interact over time.

As shown in Figure~\ref{fig:recloop}, \textsc{RecLoop} consists of two coupled modules: a recommender module and an LLM-powered user agent module.
For each user $u$, the simulation starts from an initial interaction history $\mathcal{S}{u}^{0}$.
The simulation then proceeds through multiple interaction cycles.
At cycle $t$, the recommender $f{\theta}^{t}$ takes the accumulated history $\mathcal{S}{u}^{t}$ as input and produces an exposure list of item candidates,
$
\mathcal{E}{u}^{t} = f_{\theta}^{t}(\mathcal{S}{u}^{t}).
$
The user agent evaluates the exposed candidates according to its current profile and memory, and selects one item from $\mathcal{E}{u}^{t}$ as feedback.
The selected item is appended to the user history to form $\mathcal{S}{u}^{t+1}$.
After feedback from all users is collected, the newly generated interactions are incorporated into the simulation dataset, and the recommender is retrained before the next cycle.
Through this recurrent process, each recommendation is conditioned on all prior simulated decisions, enabling  information cocoon effects to emerge and accumulate across cycles.
The internal design of the user agent, including its dynamic profile, dual memory system, reflection mechanism, and action module, is introduced in the following subsection.

\subsection{LLM-powered User Simulators for Sequential Recommendation}

To simulate user behavior in the closed loop, we model each user with an LLM-powered simulator that selects items from the exposure lists generated across simulation cycles~\cite{sukiennik2025simulating, zhang2024generative}.
Each simulator is designed to preserve user-specific preferences while allowing the simulated behavior to evolve with newly received recommendations.
As illustrated in Figure~\ref{fig:recloop}, the user agent consists of four components: (1)~a dynamic user profile that represents the user's preferences and behavioral patterns, (2)~a dual memory system that maintains both recent choices and long-term interaction history, (3)~a periodic reflection mechanism that distills accumulated experience into updated behavioral summaries, and (4)~an interaction module that takes the profile, memory, and current exposure list as input and selects an item from the recommended candidates.

\smallskip
\noindent\textbf{Dynamic User Profile.}
User preferences in sequential recommendation are generally non-stationary, as users may exhibit new interests and behavioral tendencies after repeated exposure and feedback cycles~\cite{liu2026diagnostic}.
To capture both historical consistency and simulated preference evolution, each agent maintains a base profile and a stage-wise profile.
For user $u$, the base profile $P_{u}^{\mathrm{base}}$ is constructed offline from the user's pre-simulation historical interaction records by prompting an LLM to synthesize a first-person psychological profile covering core interests, behavioral tendencies, and purchasing motivations.
This base profile provides the agent with an initial preference anchor derived from real user history.
The stage-wise profile $P_{u}^{t}$ is initialized as $P_{u}^{0}=P_{u}^{\mathrm{base}}$ and updated every $\Delta$ cycles through reflection, which will be introduced later.
Between reflection steps, $P_{u}^{t}$ remains fixed and is used for item selection.
This design allows the agent to reflect simulated preference changes while retaining its historical preference base.
An example of the generated user profile is shown in Figure~\ref{appendix:profile_b}.
The prompts used to generate $P_{u}^{\mathrm{base}}$ and update $P_{u}^{t}$ are provided in Appendix Figure~\ref{appendix:profile_a_part1} and Appendix Figure~\ref{appendix:profile_a_part2}, respectively.

\smallskip
\noindent\textbf{Dual Agent Memory.}
The user profile provides a compact summary of preferences, but it cannot preserve all interaction details needed for cycle-level decisions and long-term reflection.
Therefore, each agent maintains a two-tier memory system that records simulated interactions at different temporal scales.
Inspired by long- and short-term preference modeling in sequential recommendation~\cite{chi2022longshorttermpreferencemodeling, ijcai2018p511,10.1145/3459637.3482016, cheng2026modeling}, the memory system balances recency and completeness.
The short-term memory $M_{u}^{t}$ is a sliding window over the most recent $W$ cycles in the interaction history $\mathcal{S}{u}^{t}$.
It provides the interaction module with recent selections and rationales, helping the agent make decisions based on its current intent.
The long-term memory is represented by the complete interaction history $\mathcal{S}{u}^{t}$, which stores all simulated interactions up to cycle $t$ and provides the full behavioral trajectory for periodic reflection.

\smallskip
\noindent\textbf{Periodic Reflection.}
The long-term memory $\mathcal{H}_{u}^{t}$ preserves the complete simulated interaction history. 
However, directly relying on detailed interaction records makes it difficult for the agent to maintain a concise and coherent preference profile over extended cycles.
To convert accumulated experience into an updated profile, we introduce a periodic reflection mechanism inspired by reflective agents~\cite{park2023generative,shu2024rah,gu2025r}.

The reflection mechanism is triggered every $\Delta$ cycles.
At each reflection step, the agent analyzes the complete interaction history $\mathcal{H}{u}^{t}$ and produces a behavioral summary covering preference patterns, decision factors, and behavioral trends.
The stage-wise profile $P{u}^{t}$ is then updated using this summary while retaining the base profile $P_{u}^{\mathrm{base}}$ as the historical anchor.
Updating the profile periodically rather than at every cycle reduces computational cost and makes the update less sensitive to individual outlier interactions.
This design allows the agent to reflect preference changes accumulated during simulation while preserving the preference base derived from pre-simulation history.
The reflection prompt template is provided in Appendix Figure~\ref{appendix:reflection}.

\smallskip
\noindent\textbf{Action.}
At each cycle $t$, the agent selects one item from the exposure list $\mathcal{E}{u}^{t}$ based on its current profile and memory context.
This decision is implemented as a single LLM call that takes a structured prompt $\pi(P{u}^{t}, M_{u}^{t}, \mathcal{E}{u}^{t})$ as input and returns a structured response:
\begin{equation}
(i_{u}^{t}, r_{u}^{t}) = \mathrm{LLM}\big(\pi(P_{u}^{t}, M_{u}^{t}, \mathcal{E}{u}^{t})\big),
\end{equation}
where $i_{u}^{t} \in \mathcal{E}{u}^{t}$ denotes the selected item and $r$ denotes reason of the selection.
The prompt $\pi$ is organized into four semantic layers: a context layer that provides $P_{u}^{t}$ as the system role, a memory layer that supplies recent selections from $M_{u}^{t}$, a perception layer that presents the structured description of each candidate in $\mathcal{E}{u}^{t}$, and a constraint layer that enumerates the exact item IDs in $\mathcal{E}{u}^{t}$.
The constraint layer restricts the LLM output to valid candidates and prevents the generation of plausible but non-existent item identifiers~\cite{zhang2024generative}.
After selection, the item is appended to the interaction history $\mathcal{S}{u}^{t+1} = \mathcal{S}{u}^{t} \oplus i_{u}^{t}$, providing newly generated interaction data for recommender retraining in the next cycle.
The complete prompt template for $\pi$ is provided in Appendix Figure~\ref{appendix:decision}.

\subsection{Recommender}

Information cocoon formation is closely tied to the exposure patterns produced by recommender systems.
To examine how different recommendation paradigms influence these patterns under feedback loops, \textsc{RecLoop} places the recommender in an iterative setting where model-generated exposure is repeatedly followed by simulated user feedback and model retraining.
At each simulation cycle $t$, the recommender takes the accumulated interaction history $\mathcal{S}{u}^{t}$ of user $u$ as input, generates a ranked exposure list $\mathcal{E}{u}^{t}$, and passes it to the user agent for evaluation.

To reflect how deployed recommenders evolve with user behavior, we retrain the model at the end of every cycle using the updated interaction sequences collected from all agents.
This ensures that recommendations in later cycles are conditioned on the feedback produced within the simulation, rather than being generated by a fixed model trained only on the initial data.
As a result, \textsc{RecLoop} allows us to observe how exposure patterns generated by different recommenders accumulate and evolve over repeated feedback cycles.

\smallskip
\noindent\textbf{ID-based vs.\ Generative Recommender.}
We compare two sequential recommendation paradigms that differ in how items are represented and retrieved.
ID-based sequential models~\cite{SASRec, sun2019bert4rec, yang2024uncoveringselectivestatespace} represent each item as an atomic identifier and predict the next item by scoring candidates in a fixed item set.
Generative recommenders~\cite{tiger, deng2025onerec, MiniOneRec, 10.1145/3773771, wang2024content}, in contrast, tokenize each item into multi-level discrete sematic codes, then identify target items by autoregressively generating code sequences and finally maps code sequences to item.
This distinction changes the space in which exposure is produced: ID-based models concentrate exposure through item-level scoring, whereas generative models produce exposure through sequential decisions over a hierarchical code space.
This structural difference is central to our study of information cocoons.
Prior work has shown that ID-based recommenders can reinforce popularity and homogenize exposure through repeated feedback, as user representations and item popularity become increasingly aligned over retraining cycles~\cite{lin2025recommendation, 10.1145/3306618.3314288}.
However, it remains unclear whether generative recommenders follow the same cocoon dynamics, since their recommendations are not obtained by directly scoring all atomic item IDs.
Instead, exposure may be shaped by token-level generation, coarse-to-fine code dependencies, and decoding procedures in the learned code space.
\textsc{RecLoop} therefore evaluates both paradigms under the same closed-loop simulation protocol, allowing us to compare not only item-level exposure concentration but also how cocoon effects emerge within the generative code hierarchy.

%% file: sections/metrics.tex
\section{MEASURING INFORMATION COCOON EFFECTS}
\label{sec:metrics}

To quantify information cocoon effects in closed-loop recommendation, we use two groups of metrics.
The first group consists of exposure-level metrics, which measure cocoon effects from the item lists shown to users.
These metrics capture whether recommendations become less diverse for individual users, more similar across users, or more concentrated on a small subset of items or categories.
They provide a model-agnostic view of how the observable recommendation environment changes over simulation cycles.

Beyond these observable exposure outcomes, we further introduce code-space metrics for generative recommenders.
Unlike ID-based models that directly score atomic items, generative recommenders produce items by autoregressively generating multi-level codes.
This makes it possible to examine cocoon formation inside the learned generative space rather than only at the final item level.
Specifically, we track how generated codes are distributed across different code layers over time, which reveals where diversity is lost or preserved during the generative recommendation process.
\subsection{Exposure Diversity}
\label{sec:exposure_diversity}

We define \emph{exposure} as the candidate list $\mathcal{E}_{t,u}$ that the recommender presents to user $u$ at cycle $t$, and organize our metrics along a micro-to-macro progression: how broad each individual user's exposure is (\emph{intra-user diversity}), how similar recommendations become across users (\emph{inter-user homogenization}), how broadly the system reaches the item and category space (\emph{coverage}), and how evenly exposure is distributed among reached items (\emph{exposure concentration})~\cite{diversity,stamenkovic2021choosing,anwar2024homogenization,fleder2009blockbuster,areeb2023filter,sukiennik2024deep, bellogin2013comparative}.

\smallskip
\noindent\textbf{Intra-user Diversity.}
For a given user $u$ at simulation cycle $t$, we quantify the breadth of their recommended content using category entropy.
\emph{Category entropy at level $l$} captures distributional uniformity within a single cycle:
\begin{equation}
    E_{t,l} = -\frac{1}{|U|}\sum_{u \in U}\sum_{c \in \mathcal{C}_l} p_{t,u}(c) \log p_{t,u}(c),
    \label{eq:entropy}
\end{equation}
where $\mathcal{C}_l$ denotes the complete set of categories at hierarchy level $l$, and $p_{t,u}(c)$ is the fraction of items in $\mathcal{E}_{t,u}$ belonging to category $c$ at hierarchy level $l$, and the outer average is taken over all users.
Declining entropy over cycles indicates that per-cycle exposure is concentrating into fewer categories.

\smallskip
\noindent\textbf{Inter-user Homogenization.}
Individual narrowing alone does not necessarily imply a system-level cocoon: different users may each narrow into distinct niches~\cite{shivaram2022reducing,anwar2024homogenization}.
A collective cocoon emerges when narrowing is accompanied by convergence, i.e., different users are increasingly exposed to the same items.
We measure this effect using the average pairwise Jaccard similarity between users' exposure lists:
\begin{equation}
    \bar{J}^{t} = \frac{1}{\binom{|\mathcal{U}|}{2}} 
    \sum_{u < v}
    \frac{|\mathcal{E}_{u}^{t} \cap \mathcal{E}_{v}^{t}|}
    {|\mathcal{E}_{u}^{t} \cup \mathcal{E}_{v}^{t}|}.
    \label{eq:jaccard}
\end{equation}
A larger $\bar{J}^{t}$ indicates stronger inter-user homogenization, as different users receive more overlapping exposure.

\smallskip
\noindent\textbf{System-level Coverage.}
Entropy and Jaccard similarity characterize diversity and overlap at the user-list level, but they do not reveal how broadly the recommender explores the full item catalog across all users~\cite{vargas2011rank}.
Coverage measures this aggregate reach directly:
\begin{equation}
\mathrm{Cov}^{t} =
\frac{\left|\bigcup_{u \in \mathcal{U}} \mathcal{E}_{u}^{t}\right|}
{|\mathcal{I}|},
\qquad
\mathrm{CatCov}_{l}^{t} =
\frac{\left|\bigcup_{u \in \mathcal{U}} \mathcal{C}_{u,l}^{t}\right|}
{|\mathcal{C}_{l}|},
\label{eq:coverage}
\end{equation}
where $\mathcal{E}_{u}^{t}$ is the top-$K$ exposure list generated for user $u$ at cycle $t$ before agent selection, $\mathcal{I}$ is the full item catalog, $\mathcal{C}_{u,l}^{t}$ denotes the set of level-$l$ categories covered by $\mathcal{E}_{u}^{t}$, and $\mathcal{C}_{l}$ is the full set of level-$l$ categories.
Higher $\mathrm{Cov}^{t}$ and $\mathrm{CatCov}_{l}^{t}$ indicate broader system-level exploration of the item and category spaces, whereas lower values suggest that exposure is concentrated within a smaller portion of the catalog.

\smallskip
\noindent\textbf{Exposure Concentration.}
Coverage measures how many items are reached at the system level, but it does not capture how unevenly exposure is distributed among those items.
A recommender may cover many items while still allocating most exposure to a small subset.
We measure this imbalance using the collective Gini coefficient~\cite{fleder2009blockbuster}:
\begin{equation}
G_{\mathrm{coll}}^{t} =
\frac{
\sum_{i \in \mathcal{I}} \sum_{j \in \mathcal{I}}
\left| s_{i}^{t} - s_{j}^{t} \right|
}
{2|\mathcal{I}|^{2}\bar{s}^{t}},
\label{eq:exposure_gini}
\end{equation}
where
$
s_{i}^{t} = \sum_{u \in \mathcal{U}} \mathbf{1}[i \in \mathcal{E}_{u}^{t}]
$
counts how many users are exposed to item $i$ at cycle $t$, and $\bar{s}^{t}$ is the average exposure count over all items.
A larger $G_{\mathrm{coll}}^{t}$ indicates stronger system-level exposure concentration, meaning that impressions are increasingly dominated by a smaller subset of items.

All exposure-level metrics above are computed at each cycle to capture the instantaneous state of the recommendation environment.
In Section~\ref{sec:experiments}, we analyze their trajectories over time to distinguish transient fluctuations from sustained cocooning trends.

\subsection{Code-Space Structural Cocoon}
\label{sec:structural_cocoon}
Beyond exposure-level metrics, we introduce a code-space metric to characterize \emph{structural cocoon} effects in generative recommenders.
Exposure-level metrics measure what users receive, but they treat the item space as flat and cannot reveal how concentration emerges inside the model's learned generation space.
This limitation is especially important for generative recommendation, where items are not directly selected as atomic IDs but generated through structured code sequences.

Recent work on deep filter bubbles~\cite{sukiennik2024deep} studies hierarchical narrowing over externally defined item taxonomies, such as top-level, mid-level, and fine-grained categories.
In contrast, our metric analyzes the learned code hierarchy used by generative recommenders themselves.
For each item, a generative recommender assigns an $m$-level discrete code sequence $(c^{(0)}, c^{(1)}, \ldots, c^{(m-1)})$.
Codes at earlier levels are often associated with coarser item distinctions, while later levels provide finer-grained refinements.
Because this hierarchy is learned during item tokenization rather than imposed by external labels, it provides a model-internal view of how recommendation diversity is preserved or lost across generation levels.
We refer to concentration in this learned code hierarchy as a \emph{structural cocoon}.

\begin{definition}[Layer-wise Code Entropy]
\label{def:code_entropy}
Generative recommenders select items through multi-layer code generation rather than direct item scoring.
Therefore, two exposed items may be different at the item level but still share the same codes at certain generation layers.
If many exposed items share only a small set of codes at layer $n$, the recommender has already narrowed its generation space at that layer.
To identify where such narrowing occurs, we measure the entropy of generated codes separately for each layer.

For a generative recommender at simulation cycle $t$, the \emph{code entropy at layer $n$} is
\begin{equation}
H_{t,n} = -\frac{1}{|\mathcal{U}|}\sum_{u \in \mathcal{U}}\sum_{c \in \mathcal{V}_n} p_{t,u,n}(c) \log p_{t,u,n}(c),
\label{eq:code_entropy}
\end{equation}
where $\mathcal{V}_n$ is the code vocabulary at layer $n$, and $p_{t,u,n}(c)$ is the empirical frequency of code $c$ at layer $n$ within the exposure list $\mathcal{E}_{u}^{t}$.
Specifically, for each item in $\mathcal{E}_{u}^{t}$, we look up its layer-$n$ code from the learned item codebook and compute the frequency distribution over $\mathcal{V}_n$.
The outer average is taken over all users, so $H_{t,n}$ measures the average user-level diversity of generated codes at layer $n$.
The normalized variant $\hat{H}_{t,n} = H_{t,n} / \log |\mathcal{V}_n|$ enables comparison across layers when code vocabulary sizes differ.
\end{definition}

\begin{definition}[Top-$\kappa$ Code Concentration]
\label{def:topk_concentration}
To complement the user-level entropy above, we also measure system-level concentration in the generated code space.
Let $p_{t,n}(c)$ denote the empirical frequency of code $c$ at layer $n$ over all exposure lists ${\mathcal{E}_{u}^{t}}_{u \in \mathcal{U}}$.
The \emph{top-$\kappa$ concentration at layer $n$} is the cumulative probability mass of the $\kappa$ most frequent codes:
\begin{equation}
\mathrm{Top}\text{-}\kappa_{t,n} = \sum_{k=0}^{\kappa-1} p_{t,n}(c_{(k)}),
\label{eq:topk}
\end{equation}
where $c_{(0)}, c_{(1)}, \ldots, c_{(\kappa-1)}$ are codes ranked by descending probability under $p_{t,n}$.
Higher values indicate that a small number of codes dominate the system-level generation distribution at that layer.
\end{definition}

All code-space metrics are computed on exposure lists before user selection, so they characterize the recommender's generated exposure rather than the agent's final feedback.
Together, layer-wise code entropy and top-$\kappa$ code concentration capture complementary aspects of code-space narrowing: the former measures the average diversity of codes within each user's exposure list, while the latter measures whether system-wide exposure mass is dominated by a small set of codes.

Based on these two metrics, we use \emph{structural cocoon} to describe a model-level narrowing phenomenon in generative recommenders.
A structural cocoon occurs when generated exposure becomes increasingly concentrated within the learned code hierarchy, especially when this concentration is unevenly distributed across layers.
To summarize the degree of layer-wise narrowing, we report the relative entropy reduction:
\begin{equation}
\delta_n = \frac{H_{1,n} - H_{T,n}}{H_{1,n}}, \quad n = 0, 1, \ldots, m-1,
\label{eq:entropy_reduction}
\end{equation}
where layer indices run from coarsest ($0$) to finest ($m-1$).
A larger $\delta_n$ indicates a stronger reduction in average user-level code diversity at layer $n$.

The per-layer entropy reduction $\delta_n$ shows how cocoon effects distribute across code granularities.
For example, a profile such as $(\delta_0, \delta_1, \delta_2) = (0.54, 0.19, 0.13)$ indicates that the coarsest layer collapses most severely while the finer layers retain more diversity.
The semantic categories captured by discrete codes need not align with the externally defined item taxonomy used in the exposure-level metrics, so the model's code-level distribution may narrow while category-level exposure metrics remain stable, or vice versa.
Tracking $\delta_n$ therefore provides a model-level diagnostic for cocoon formation that is specific to generative recommendation architectures.

%% file: sections/experiments.tex
\section{Results and Findings}
\label{sec:experiments}

We evaluate whether and how generative recommendation changes the formation of information cocoons under closed-loop feedback.
The experiments are organized around four research questions that progress from observable exposure outcomes to the internal mechanisms and influencing factors of generative recommendation.

\textbf{RQ1: Do generative recommenders deepen information cocoons under feedback loops?}
We first compare generative recommenders with traditional sequential recommenders using exposure diversity, inter-user homogenization, coverage, and exposure concentration.

\textbf{RQ2: Where does cocoon formation emerge in the generative code space?}
We then examine whether diversity loss is uniformly distributed across code layers or concentrated at specific levels of the hierarchical code sequence.

\textbf{RQ3: How does item tokenization affect cocoon formation?}
Because generative recommenders rely on discrete item codes, we further ask whether Semantic IDs and Collaborative IDs induce different cocoon dynamics when the quantization architecture is held fixed.

\textbf{RQ4: Can model scale buffer structural cocoon formation?}
Finally, we analyze whether increasing model capacity preserves broader active code usage under repeated feedback, mitigating the collapse of semantic regions.

Together, these questions correspond to the main findings of this section: generative recommenders slow down category-level diversity loss and inter-user exposure homogenization compared with traditional sequential recommenders; cocoon effects concentrate unevenly in the generative code hierarchy; tokenization signals modulate cocoon severity; and larger models preserve broader code-space diversity.

\subsection{Simulation Setup}
\label{sec:setup}

\subsubsection{Datasets}

We conduct experiments on two Amazon product review datasets\footnote{\url{https://cseweb.ucsd.edu/~jmcauley/datasets/amazon/links.html}}~\cite{mcauley2015image}: \textbf{Office Products} and \textbf{Toys and Games}.
We select these two domains because each item carries a multi-level category hierarchy (up to four levels) shown in Figure~\ref{appendix:item_description}, which directly enables our diversity analysis across semantic granularities.
We follow the standard 5-core preprocessing protocol~\cite{tiger}, retaining users and items with at least five interactions, and order each user's interactions chronologically.
Dataset statistics after preprocessing are summarized in Table~\ref{tab:datasets}.

\begin{table}[t]
    \centering
    \caption{Dataset statistics after preprocessing. Avg.\ Len.\ is the average number of interactions per user. Categories is the number of item categories at each hierarchy level (L1/L2/L3/L4).}
    \label{tab:datasets}
    \small
    \begin{tabular}{lrrrrr}
    \toprule
    Dataset & \#Users & \#Items & \#Interactions & Avg.\ Len. & \#Categories \\
    \midrule
    Office Products & 4,905 & 2,420 & 53,258& 10.86 & (5/11/44/152) \\
    Toys \& Games   & 19,412 & 11,924 & 167,597 & 8.63 & (11/41/191) \\
    \bottomrule
    \end{tabular}
\end{table}

\subsubsection{Models}

We compare four representative recommendation models spanning two paradigms.

\smallskip
\noindent\emph{Traditional sequential models.}
\textbf{SASRec}~\cite{SASRec} and \textbf{Mamba4Rec}~\cite{liu2024mamba4recefficientsequentialrecommendation} encode items as continuous embeddings and rank candidates via inner-product scoring.

\smallskip
\noindent\emph{Generative models.}
\textbf{TIGER}~\cite{tiger} tokenizes items into semantic IDs via RQ-VAE and generates item codes autoregressively with a Transformer backbone.
\textbf{OneRec}~\cite{deng2025onerec} is a large-language-model-based generative recommender. We use the 0.5B, 1.5B, and 3B parameter scales to analyze the effect of model capacity. 
We use the MiniOneRec~\cite{MiniOneRec} codebase with supervised fine-tuning only (no reinforcement learning), so that the training objective remains comparable to the cross-entropy supervision used by the sequential baselines and any observed diversity differences can be attributed to architecture rather than training procedure.



\subsubsection{Simulation Protocol}

All models are evaluated under the closed-loop simulation framework described in Section~\ref{sec:framework}.
Each simulation runs for $T=15$ cycles over the complete user set of each dataset.
Each user is initialized with an interaction history of maximum length 50, drawn from the user's earliest historical records, so that the simulated trajectory continues from a realistic starting point rather than from a cold start.
At each cycle, the recommender generates a top-$K$ candidate list with $K=5$ for each user, the LLM-based agent (Qwen3-8B with temperature $=0$) returns deterministic feedback, and the model is retrained on the updated interaction history.
The agent's short-term memory window is set to $W=5$ cycles, and the reflection mechanism is triggered every $\Delta=5$ cycles.
All models share the same user set, initial interaction histories, and simulation parameters to ensure controlled comparison.



\subsubsection{Implementation Details}

All experiments are conducted on NVIDIA A800 80GB.
Recommender model retraining at each cycle adopts the hyperparameters reported in the original implementations of each backbone.
The user agent is implemented with Qwen3-8B\footnote{\url{https://huggingface.co/Qwen/Qwen3-8B}} served locally via vLLM.
Code will be released upon publication.

\subsection{RQ1: Do Generative Recommenders Deepen Information Cocoons?}
\label{sec:overall_comparison}

\begin{insightbox}
\smallskip
\noindent\textbf{Finding 1.}
Closed-loop feedback induces information cocoons across all models, but generative recommenders generally show weaker category-level narrowing and cross-user convergence than traditional sequential recommenders.
The key difference is that generative recommenders decode over hierarchical item codes, while traditional models retrieve items through atomic embedding-based matching.
\smallskip
\end{insightbox}
\smallskip
\noindent\textbf{Experimental setup.}
This research question compares two recommendation paradigms under the same closed-loop simulation protocol.
Traditional sequential recommenders, represented by SASRec and Mamba4Rec, encode user histories into continuous representations and rank candidate items through embedding-based matching over atomic item IDs.
Generative recommenders, represented by TIGER and OneRec, replace atomic item scoring with autoregressive generation over discrete item codes, where each item is represented as a multi-layer code sequence.
We use two representative models from each paradigm to analyze paradigm-level cocoon dynamics rather than to benchmark recommendation accuracy.
This design allows us to examine whether closed-loop cocoon dynamics differ between atomic item matching and hierarchical code generation.

We evaluate this question using four complementary exposure-level metrics is described in Section \ref{sec:exposure_diversity}.
Category entropy measures whether each user's accessible information range becomes semantically narrower.
Inter-user Jaccard similarity measures whether different users are pushed toward increasingly similar exposure lists.
Coverage measures how much of the item and category space remains reachable during simulation.
The Gini coefficient measures whether exposure traffic concentrates on a small number of head items.
Together, these metrics distinguish two aspects of cocoon formation: \emph{semantic diversity collapse}, where exposure becomes category-wise narrower, and \emph{head-item concentration}, where recommendation mass accumulates on a small subset of items.

\begin{figure*}[t]
    \centering
    \subfloat[Office, Level~2]{\includegraphics[width=0.19\textwidth]{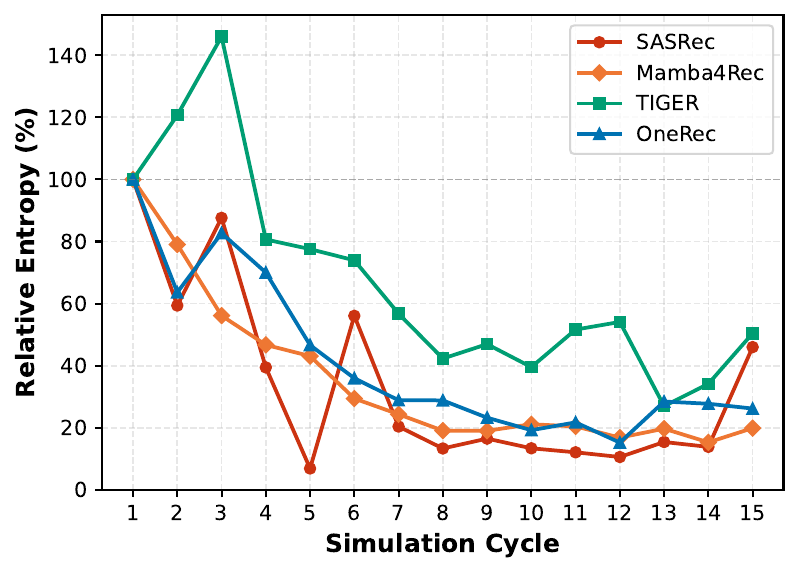}\label{fig:entropy_office_l2}}
    \hfill
    \subfloat[Office, Level~3]{\includegraphics[width=0.19\textwidth]{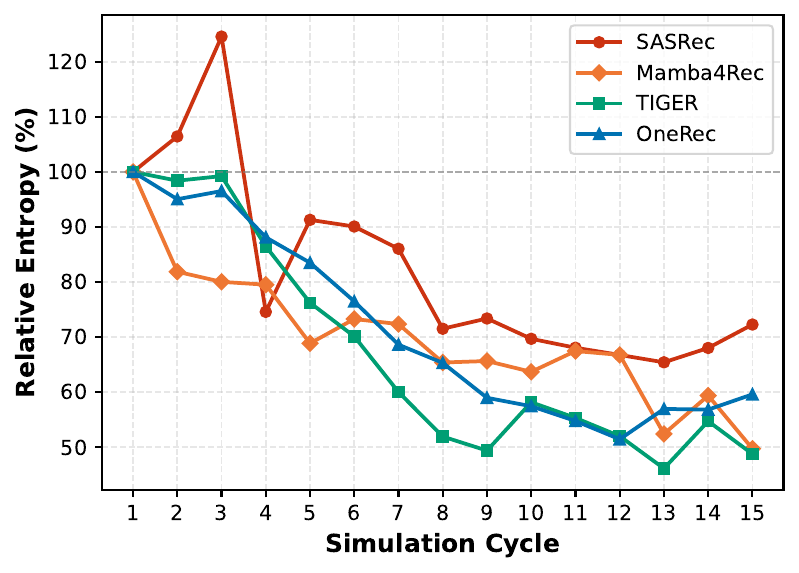}\label{fig:entropy_office_l3}}
    \hfill
    \subfloat[Office, Level~4]{\includegraphics[width=0.19\textwidth]{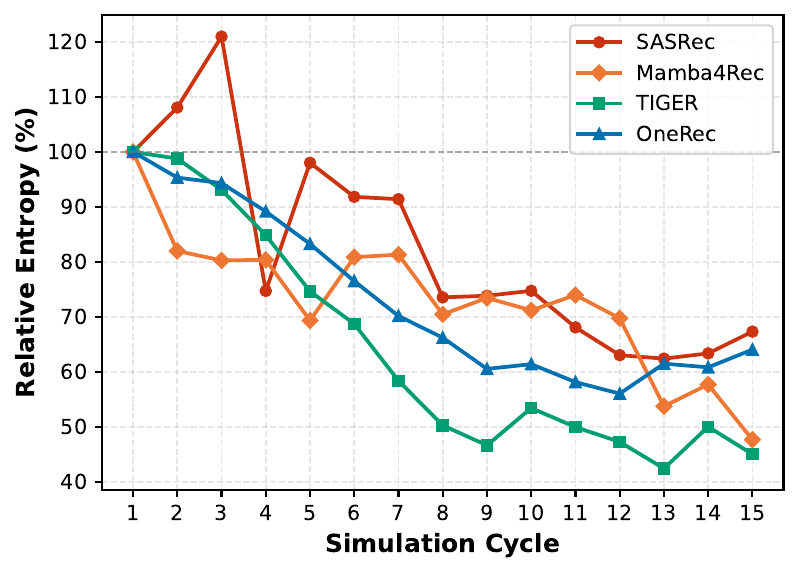}\label{fig:entropy_office_l4}}
    \hfill
    \subfloat[Toys, Level~2]{\includegraphics[width=0.19\textwidth]{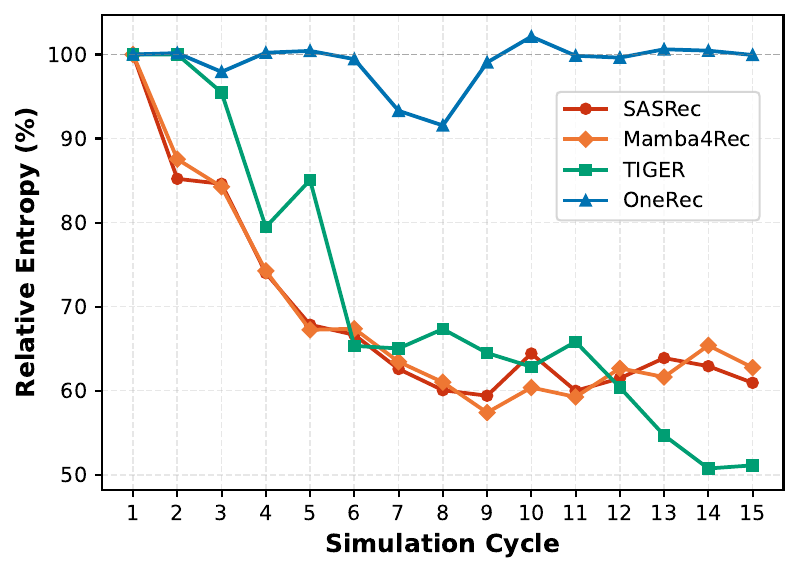}\label{fig:entropy_toys_l2}}
    \hfill
    \subfloat[Toys, Level~3]{\includegraphics[width=0.19\textwidth]{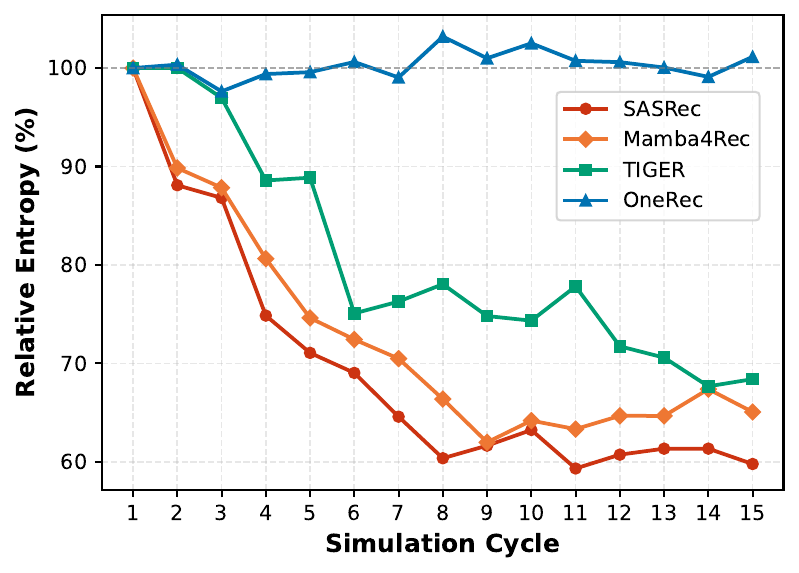}\label{fig:entropy_toys_l3}}
    \caption{Category entropy normalized to Cycle~1 across 15 simulation cycles for Office Products (Levels~2--4) and Toys and Games (Levels~2--3). Across panels, SASRec and Mamba4Rec show steep early entropy loss, TIGER declines more unevenly with occasional spikes, and OneRec also trends downward but with more oscillatory mid-cycle rebounds.}
    \label{fig:entropy_comparison}
\end{figure*}

\smallskip
\noindent\textbf{Category entropy: generative models slow down semantic narrowing.}
Figure~\ref{fig:entropy_comparison} shows that repeated feedback reduces category entropy in most model--dataset--granularity combinations.
This confirms that closed-loop recommendation generally narrows the semantic range of user exposure over time.
However, the rate and stability of entropy loss differ substantially across paradigms.
SASRec and Mamba4Rec show the sharpest early drops, indicating that embedding-based matching quickly concentrates exposure around a smaller set of categories once user feedback is repeatedly folded back into training.
TIGER declines more unevenly, with short-term spikes in several panels, suggesting that code generation does not collapse exposure as smoothly as direct item matching.
OneRec shows even stronger oscillatory behavior; on Toys and Games Level~3, its normalized entropy stays above the Cycle~1 baseline throughout all 15 cycles, ranging from 1.49 to 1.57.
Except for this case, the trajectories end below their Cycle~1 entropy, showing that generative recommendation does not remove semantic narrowing.
Instead, it slows and destabilizes the collapse process by decoding over a shared discrete code space rather than directly retrieving nearest items from an atomic embedding space.

\begin{figure*}[t]
    \centering
    \subfloat[Office Products]{\includegraphics[width=0.48\textwidth]{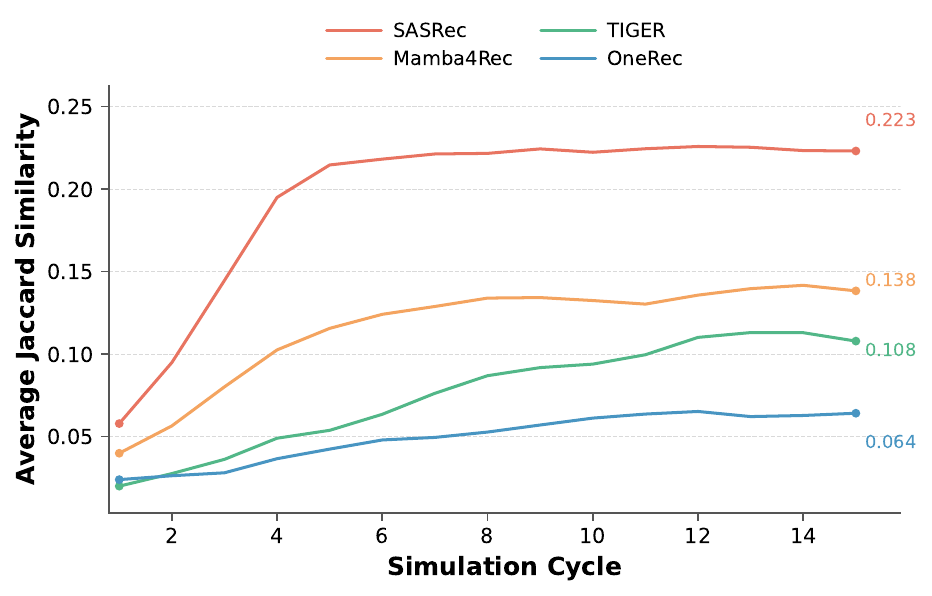}\label{fig:jaccard_office}}
    \hfill
    \subfloat[Toys and Games]{\includegraphics[width=0.48\textwidth]{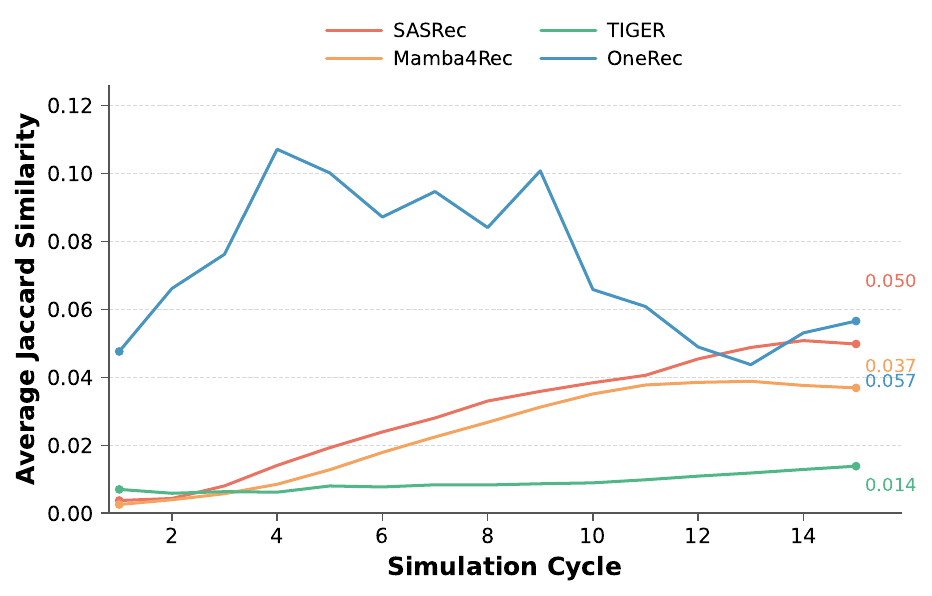}\label{fig:jaccard_toys}}
    \caption{Average pairwise Jaccard similarity across 15 simulation cycles on Office Products (left) and Toys and Games (right). Each line shows one model, where higher values indicate greater homogenization across users, with final-cycle values annotated on the right.}
    \label{fig:jaccard}
\end{figure*}

\smallskip
\noindent\textbf{Inter-user homogenization: generative models reduce cross-user convergence.}
Figure~\ref{fig:jaccard} reports the average pairwise Jaccard similarity between users' exposure lists.
On Office Products, the homogenization pattern is consistent with the entropy results.
By Cycle~15, SASRec reaches a Jaccard similarity of 0.223, Mamba4Rec reaches 0.138, TIGER reaches 0.108, and OneRec remains lowest at 0.064.
This ranking shows that traditional sequential models push different users toward increasingly similar recommendation lists, whereas generative models better preserve user-level differentiation.
A plausible mechanism is that embedding-based models repeatedly strengthen the same high-scoring item regions, causing multiple users to converge on overlapping head items.
In contrast, generative models route users through hierarchical code sequences, so different users may still arrive at distinct item groups even when their feedback histories become more concentrated.

On Toys and Games, the trend is less monotonic.
OneRec starts with a relatively high initial Jaccard similarity around 0.048, spikes to 0.103 at Cycle~4, and then partially recovers to 0.057 by Cycle~15.
This temporary spike suggests that early feedback can still concentrate a large user population into overlapping code regions.
However, the subsequent recovery indicates that the generative code space may redistribute exposure after the initial reinforcement phase.
TIGER ends with the lowest homogenization on this dataset at 0.014, while SASRec and Mamba4Rec end at 0.050 and 0.037, respectively.
Thus, generative recommenders are not uniformly better at every cycle, but they are less likely to produce steadily increasing cross-user convergence.

\begin{figure*}[t]
    \centering
    \subfloat[Office Products, Item]{\includegraphics[width=0.24\textwidth]{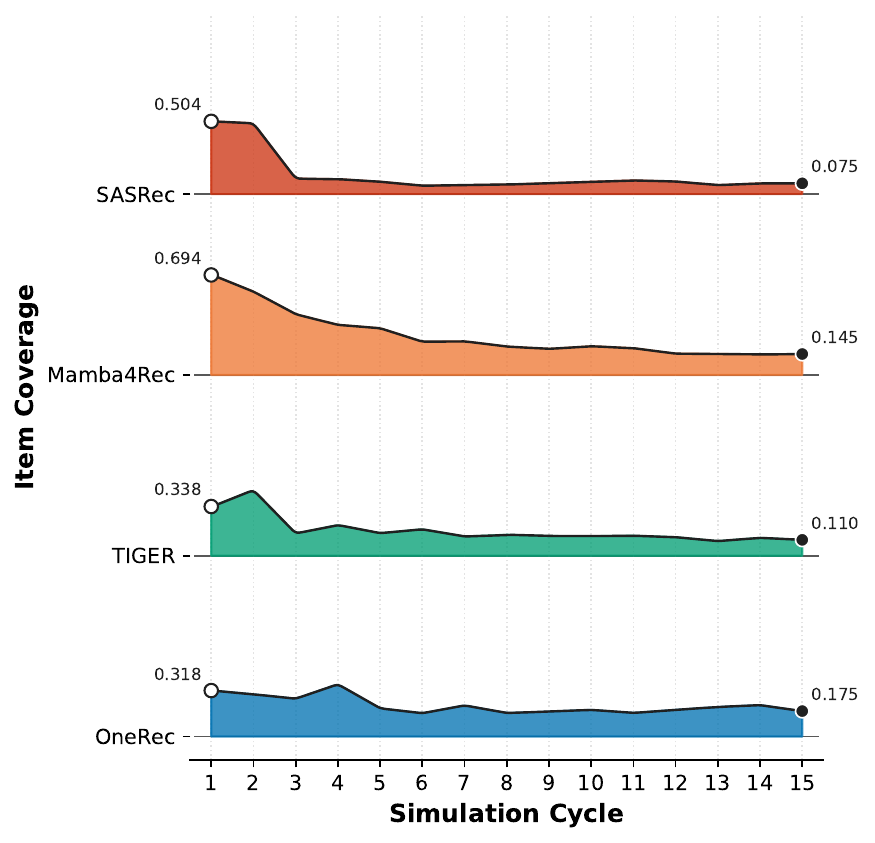}\label{fig:cov_office_item}}
    \hfill
    \subfloat[Office Products, Level~2]{\includegraphics[width=0.24\textwidth]{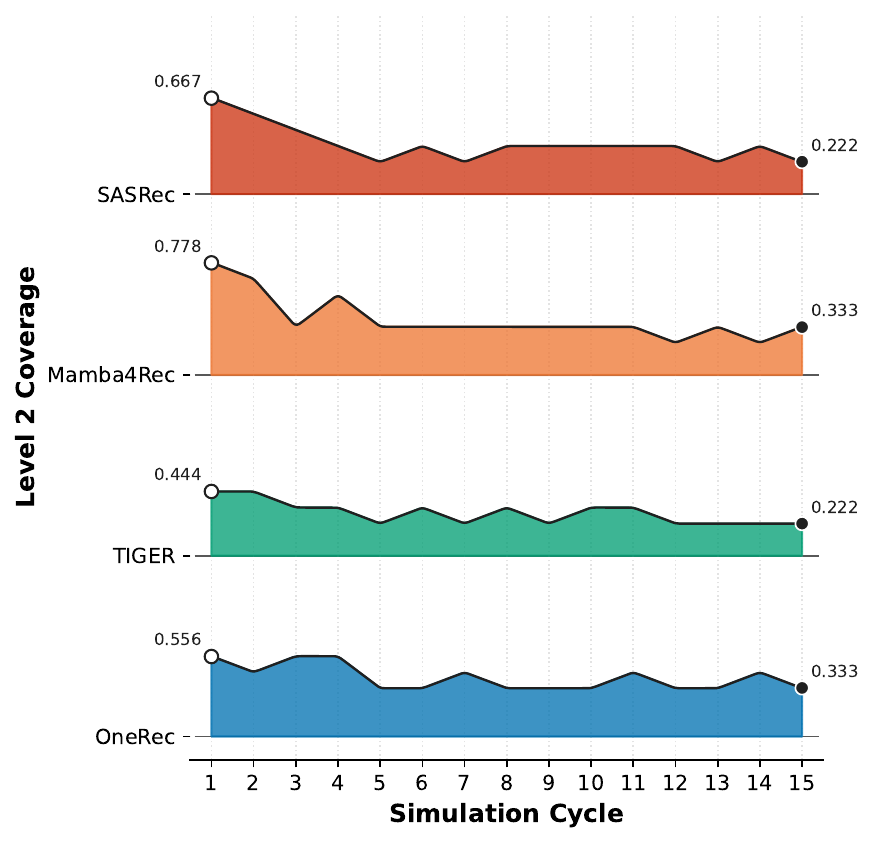}\label{fig:cov_office_l2}}
    \hfill
    \subfloat[Office Products, Level~3]{\includegraphics[width=0.24\textwidth]{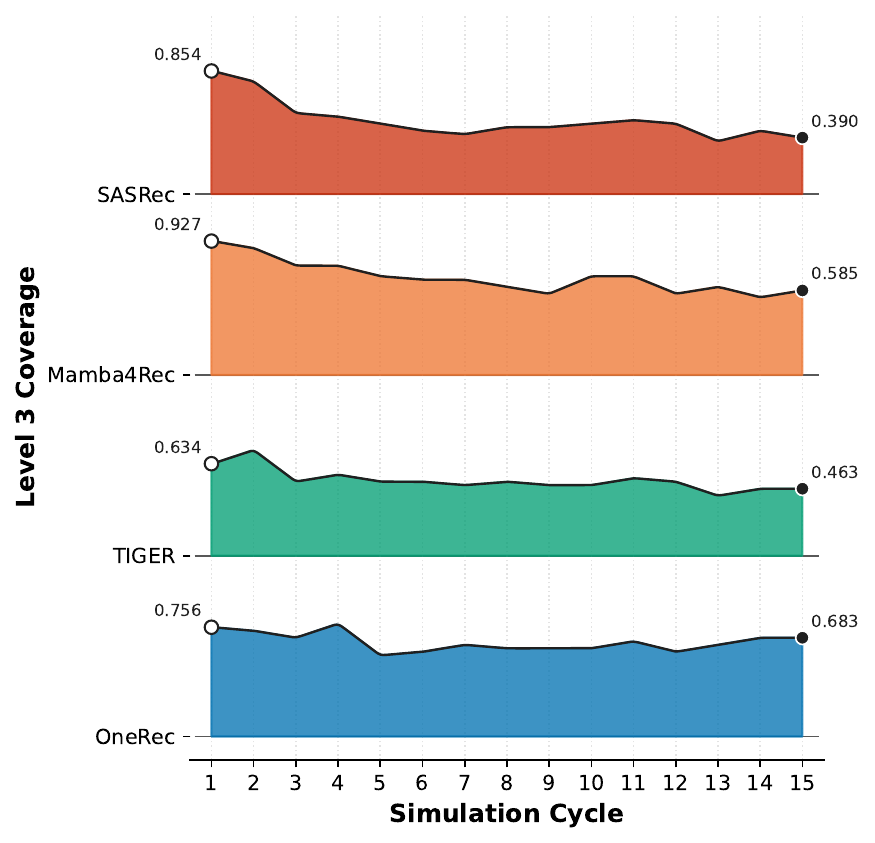}\label{fig:cov_office_l3}}
    \hfill
    \subfloat[Office Products, Level~4]{\includegraphics[width=0.24\textwidth]{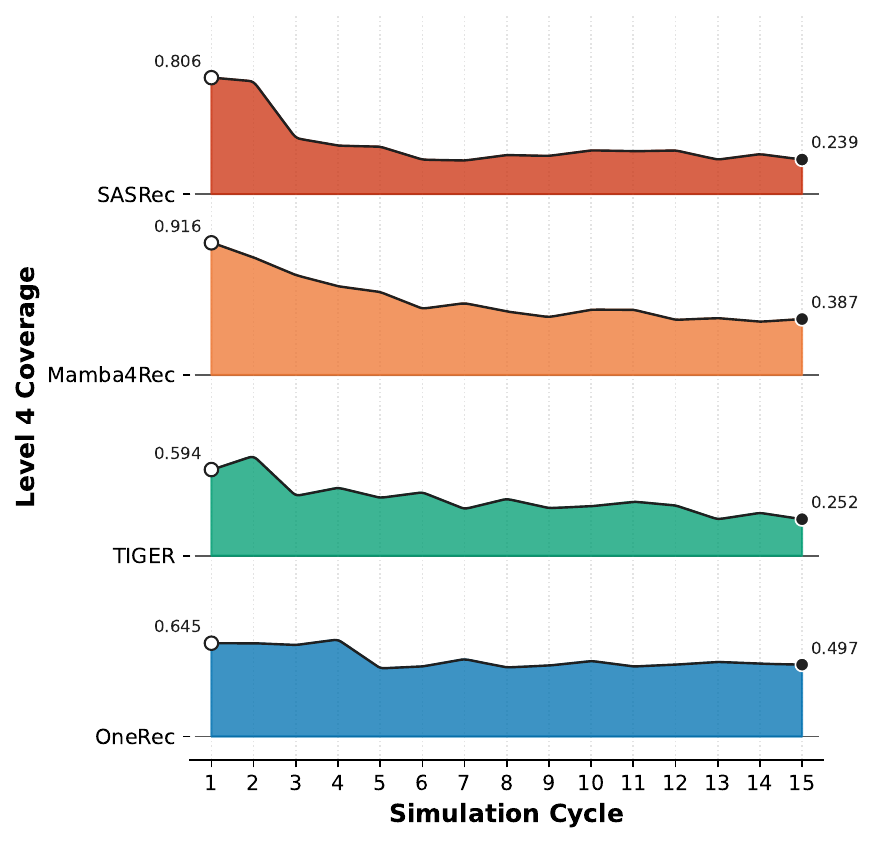}\label{fig:cov_office_l4}}
    \\
    \hfill
    \subfloat[Toys~\&~Games, Item]{\includegraphics[width=0.24\textwidth]{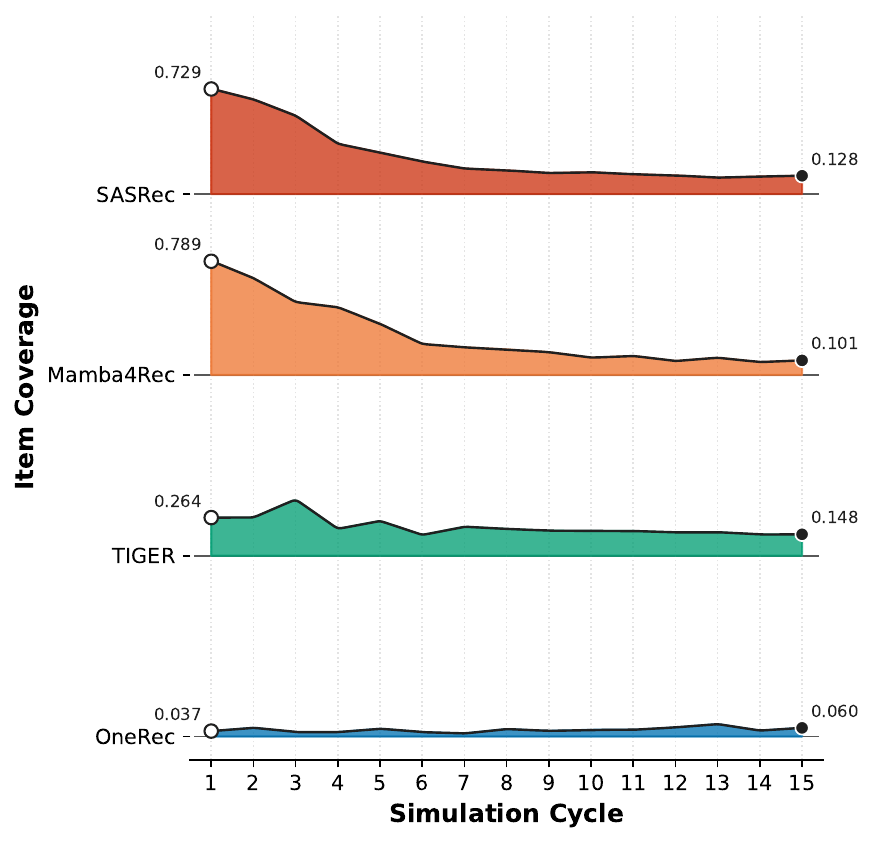}\label{fig:cov_toys_item}}
    \hfill
    \subfloat[Toys~\&~Games, Level~2]{\includegraphics[width=0.24\textwidth]{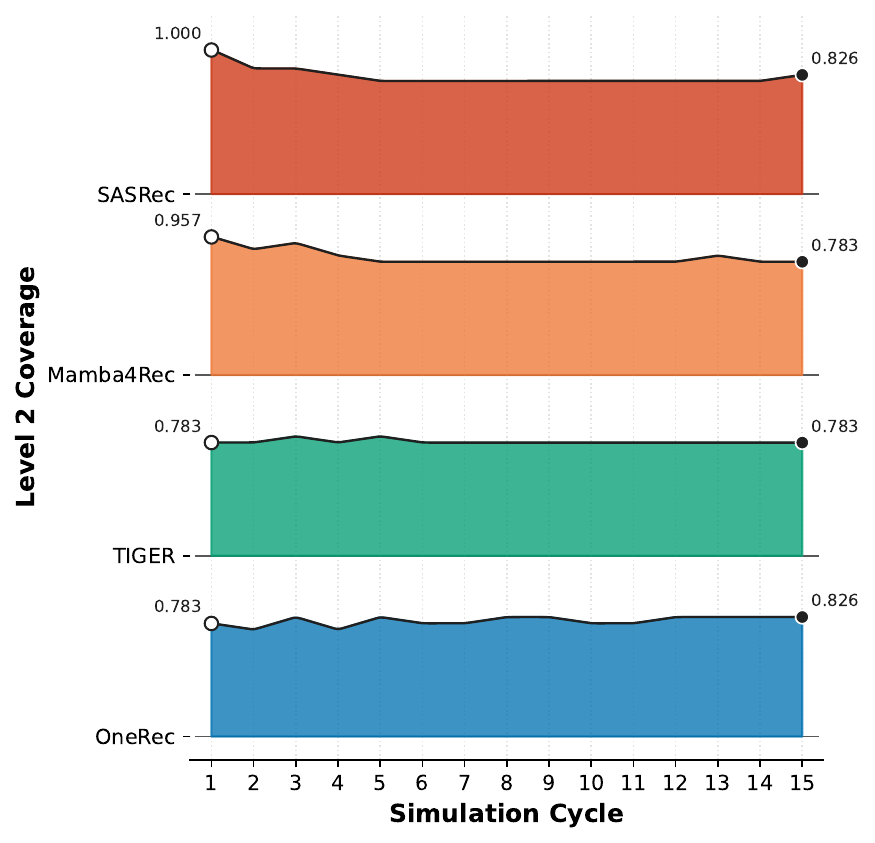}\label{fig:cov_toys_l2}}
    \hfill
    \subfloat[Toys~\&~Games, Level~3]{\includegraphics[width=0.24\textwidth]{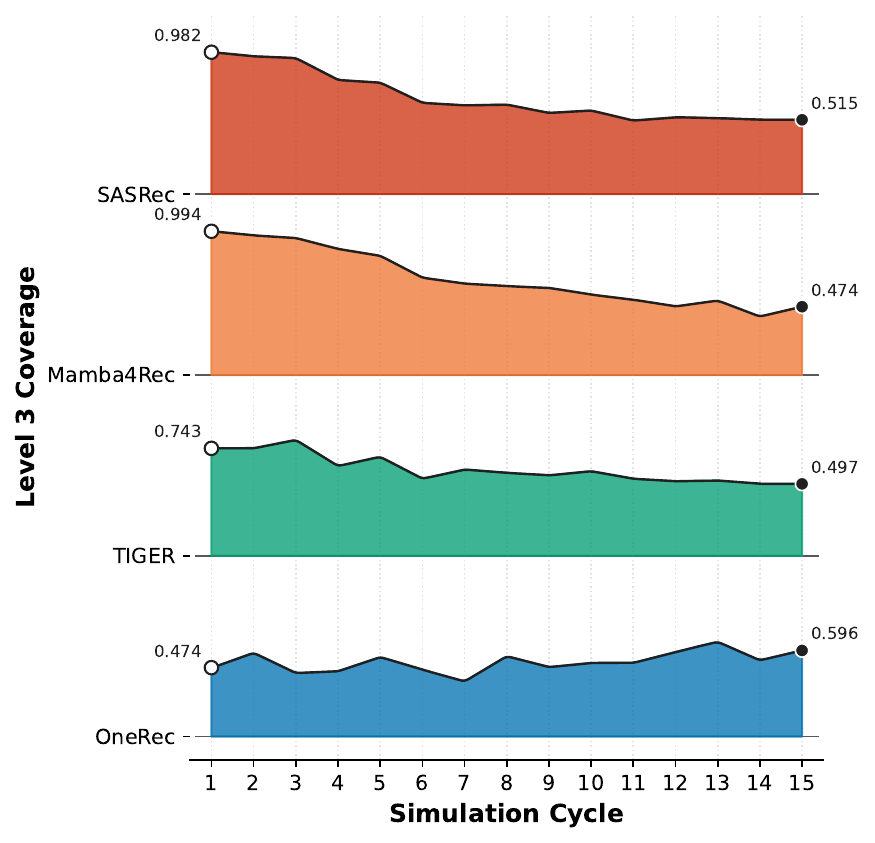}\label{fig:cov_toys_l3}}
    \hfill\null
    \caption{Item coverage and category coverage at three hierarchy levels across 15 simulation cycles. The top row presents Office Products (a--d); the bottom row presents Toys and Games (e--g). Each panel shows ridgeline trajectories for all four models from Cycle~1 to Cycle~15, with the start and end values annotated.}
    \label{fig:coverage_comparison}
\end{figure*}

\smallskip
\noindent\textbf{Coverage: generative models preserve category reachability more consistently than item reachability.}
Figure~\ref{fig:coverage_comparison} further examines whether the narrowing observed in entropy and Jaccard also reduces the reachable item and category space.
On Office Products, the four models separate clearly.
SASRec contracts fastest at every level, Mamba4Rec starts from broader exposure but reaches similarly narrow coverage by the middle cycles, TIGER slows the contraction, and OneRec retains the highest item coverage and Level-3/4 category coverage through the final cycles.
This indicates that the generative models keep a broader portion of the semantic space reachable under repeated feedback, especially at finer category granularities.

On Toys and Games, the coverage result is more dataset-dependent.
Level~2 coverage remains nearly flat for all models, suggesting that the top-level taxonomy is broad enough that even concentrated recommendation lists still span most coarse categories.
The paradigm difference becomes more visible at finer levels.
OneRec has lower absolute item coverage than the other models, which may appear inconsistent with its stronger entropy behavior.
The two observations capture different aspects of exposure.
A model can recommend from a narrower item set while still covering semantically diverse fine-grained categories; conversely, a model can expose more distinct items while concentrating them within fewer semantic regions.
Thus, coverage supports the same overall conclusion as entropy: generative recommendation is not always broader at the item level, but it tends to better preserve category-level diversity.

\begin{figure*}[t]
    \centering
    \subfloat[Toys and Games]{\includegraphics[width=0.48\textwidth]{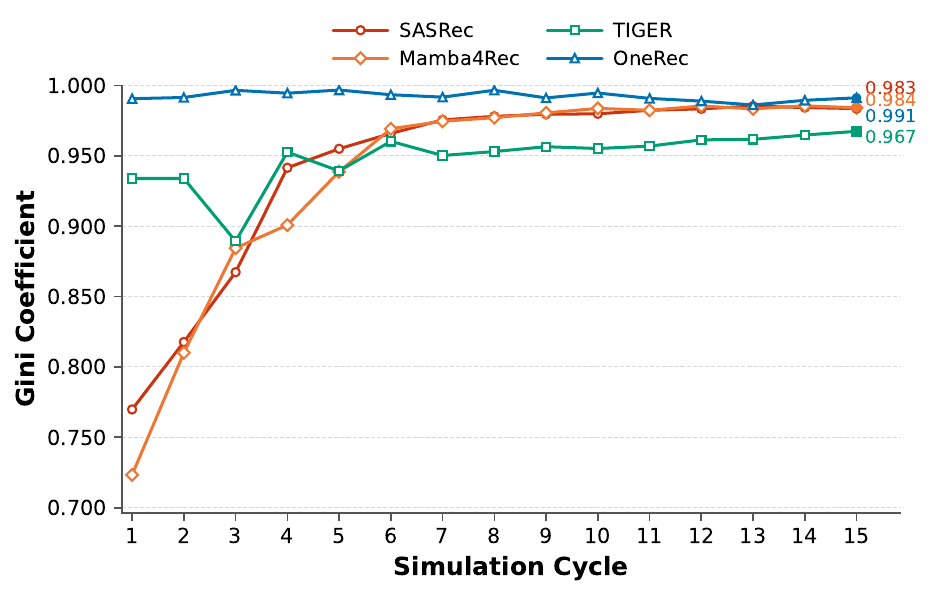}\label{fig:gini_toys}}
    \hfill
    \subfloat[Office Products]{\includegraphics[width=0.48\textwidth]{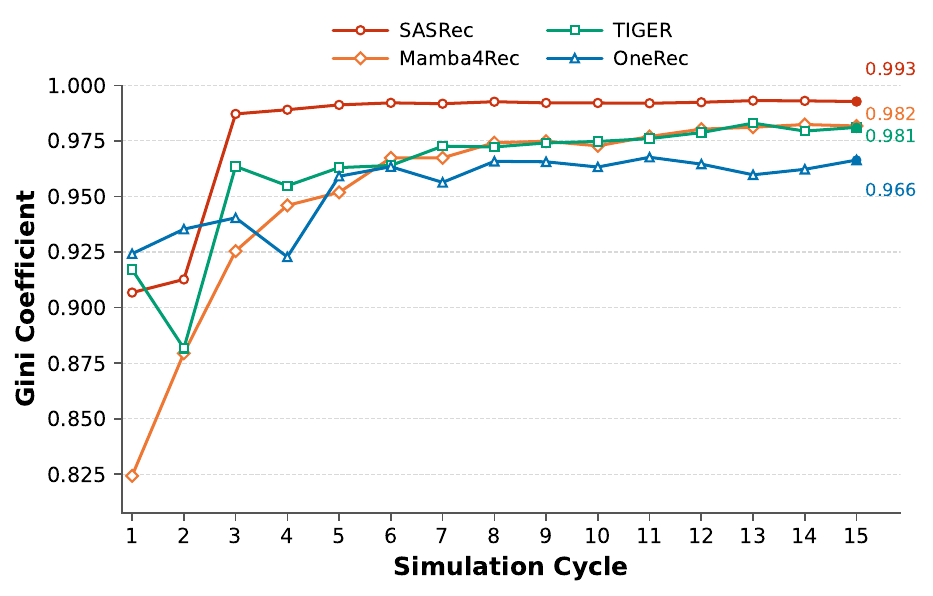}\label{fig:gini_office}}
    \caption{Gini coefficient of item exposure over 15 closed-loop simulation cycles on Toys and Games (left) and Office Products (right). All four models exhibit a generally increasing Gini, indicating progressive concentration of recommendations on a small set of popular items under closed-loop feedback.}
    \label{fig:gini}
\end{figure*}

\smallskip
\noindent\textbf{Exposure concentration: closed-loop feedback still reinforces head items.}
Figure~\ref{fig:gini} shows the Gini coefficient of item exposure across 15 simulation cycles.
Unlike entropy and Jaccard, which show clearer paradigm-level differences, the Gini coefficient increases generally across both datasets and all four models.
This universal increase indicates that closed-loop feedback systematically concentrates recommendation traffic on a smaller set of head items, regardless of whether the model uses embedding-based matching or generative code decoding.

The strength of this concentration varies by dataset and model.
On Toys and Games, TIGER starts from a highly concentrated distribution, increasing from 0.933 to 0.970, while SASRec shows the steepest absolute rise, from 0.598 to 0.907.
Mamba4Rec increases from 0.569 to 0.760 and remains the least concentrated at Cycle~15, whereas OneRec changes more mildly from 0.903 to 0.916.
On Office Products, all models converge to a narrow high-concentration band by Cycle~15, ranging from 0.966 to 0.993, with OneRec ending lowest and SASRec ending highest.
These results show that generative recommenders can slow semantic narrowing and inter-user convergence, but they do not fully prevent exposure mass from accumulating on popular items.

Overall, RQ1 shows that generative recommendation does not eliminate information cocoons under closed-loop feedback.
Instead, it mitigates two major symptoms of cocoon formation: category-level diversity loss and inter-user exposure homogenization.
Traditional sequential models tend to collapse through direct reinforcement in an atomic item-embedding space, while generative models preserve more semantic variation through hierarchical code decoding.
At the same time, the increasing Gini coefficient across all models shows that closed-loop feedback still amplifies head-item concentration.
This motivates the next question: if generative recommenders resist aggregate diversity collapse more effectively, where does cocoon formation appear inside their generated code space?

\subsection{RQ2: Where Does Cocoon Formation Emerge in the Generative Code Space?}
\label{sec:structural}

\begin{insightbox}
\smallskip
\noindent\textbf{Finding 2.}
Generative recommenders exhibit a structural cocoon in code space: diversity loss is concentrated in the coarse code layers rather than uniformly distributed across the generated code sequence.
This pattern arises because autoregressive decoding makes early tokens act as routing decisions over broad semantic regions, while later tokens still preserve finer-grained variation within the selected branches.
\end{insightbox}
\smallskip
\noindent\textbf{Experimental setup.}
This research question examines how cocoon formation emerges within the hierarchical code space of generative recommendation.
Since TIGER and OneRec represent items using multi-layer discrete codes, we can track diversity dynamics at different levels of the code hierarchy.
We therefore measure layer-wise code entropy across simulation cycles and analyze how the relative entropy reduction $\delta_k$ varies across code layers, following the structural cocoon framework introduced in Section~\ref{sec:structural_cocoon}.

\begin{figure*}[t]
    \centering
    \subfloat[Layer 0 (coarsest)]{\includegraphics[width=0.32\textwidth]{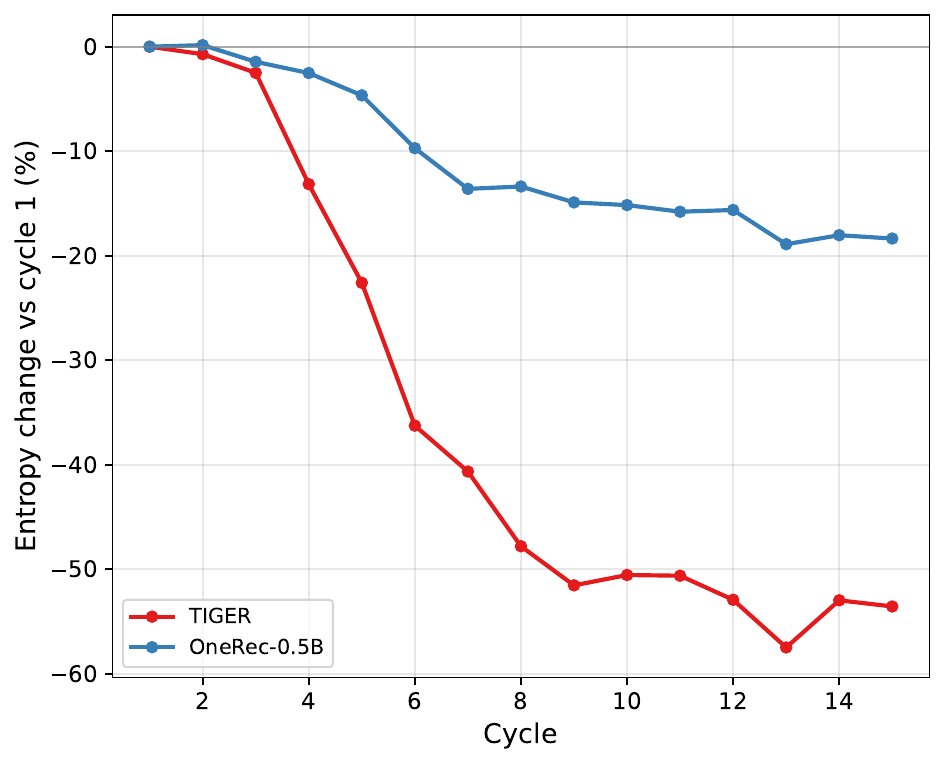}
        \label{fig:layer_entropy_l0}}
    \hfill
    \subfloat[Layer 1]{\includegraphics[width=0.32\textwidth]{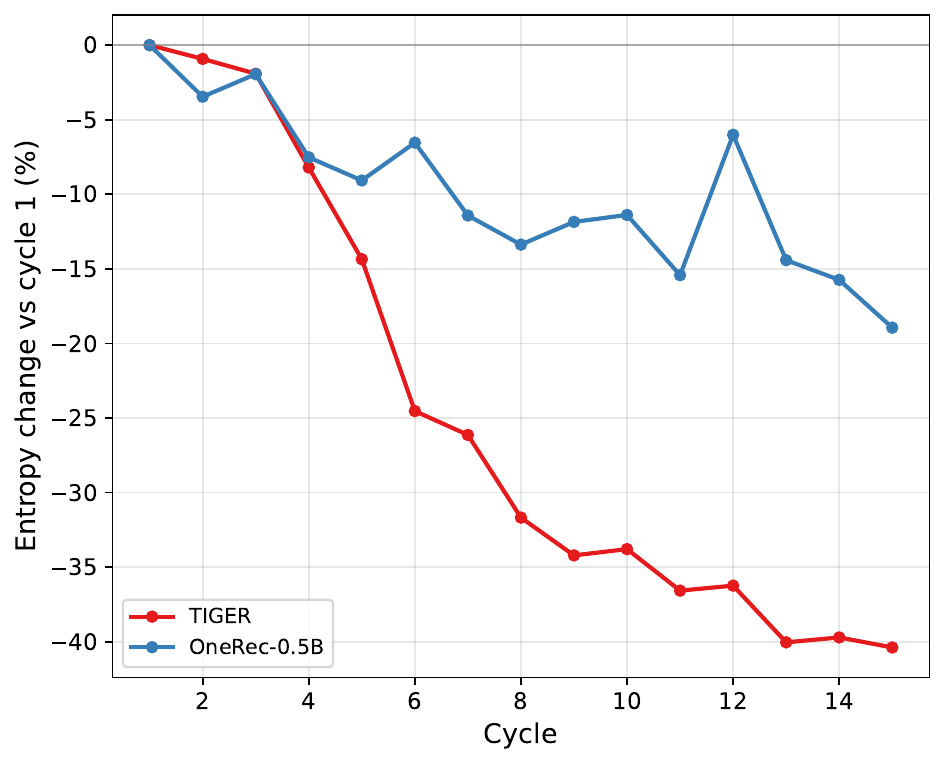}
        \label{fig:layer_entropy_l1}}
    \hfill
    \subfloat[Layer 2 (finest)]{\includegraphics[width=0.32\textwidth]{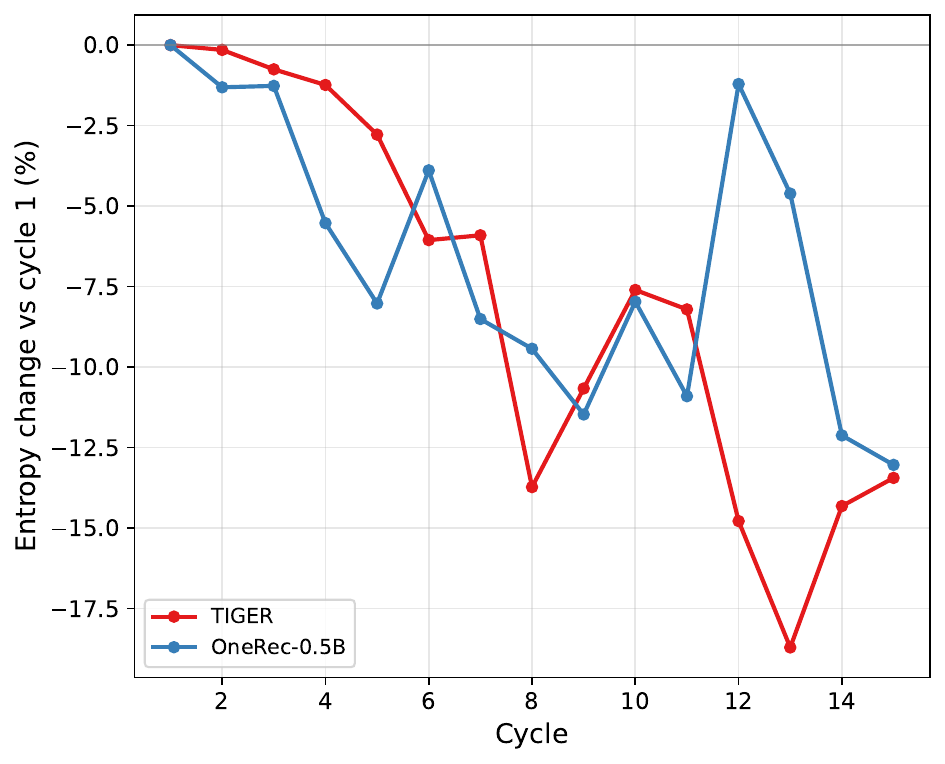}
        \label{fig:layer_entropy_l2}}
    \caption{Layer-wise code entropy across simulation cycles for TIGER and OneRec on the Office Products dataset. Each subfigure reports the relative entropy change (\%) at one code layer. Finer-grained layers undergo less entropy reduction across simulation cycles compared to coarser layers, revealing a structural cocoon formation.}
    \label{fig:layer_entropy}
\end{figure*}

\smallskip
\noindent\textbf{Layer-wise entropy dynamics.}
Figure~\ref{fig:layer_entropy} shows that cocoon severity is unevenly distributed across the generated code hierarchy.
For TIGER on Office Products, entropy reduction from Cycle~1 to Cycle~15 reaches 53.6\% at Layer~0, 40.4\% at Layer~1, and only 13.4\% at Layer~2.
OneRec shows a milder but similar pattern, with entropy reduction of 18.4\% at Layer~0, 18.9\% at Layer~1, and 13.4\% at Layer~2.
Across both models, the finest layer remains more stable than the coarse layers, indicating that diversity loss does not occur uniformly over the entire code sequence.
Instead, the feedback loop first narrows the set of coarse code regions before substantially reducing fine-grained variation.

\smallskip
\noindent\textbf{Structural cocoon pattern.}
This layer-wise asymmetry reveals a structural form of cocoon formation that is specific to generative recommendation.
In autoregressive code generation, the early code layers determine broad routing decisions: once a coarse code is selected, subsequent tokens are generated within a more restricted branch of the code hierarchy.
As feedback cycles repeat, user exposure can become concentrated around a smaller set of coarse code regions, even while the model continues to generate diverse fine-layer codes inside those regions.
Thus, the model does not simply collapse to a few individual items; rather, it funnels users into narrower semantic regions while preserving local diversity within those regions.
This explains why generative recommenders can slow aggregate exposure collapse in RQ1 while still forming a cocoon internally in code space.

Traditional sequential models do not expose an analogous layer-wise code sequence, so this structural pattern cannot be directly measured for them.
Their cocoon effect is observed mainly through exposure-level metrics such as category entropy, Jaccard similarity, coverage, and Gini concentration.
By contrast, generative recommenders allow us to locate the cocoon formation process inside the model's discrete generation space.

\smallskip
\noindent\textbf{First-token concentration.}
To further quantify the coarse-level bottleneck, we examine the entropy and top-$k$ concentration of the first generated token, i.e., the Layer~0 code.
This token is important because it acts as the first routing decision in the generated code sequence.
For TIGER on Office Products, normalized first-token entropy drops from 0.895 at Cycle~1 to 0.451 at Cycle~15, corresponding to a 49.6\% relative decrease.
Meanwhile, the cumulative probability mass of the top-10 Layer~0 codes rises from 27.0\% to 86.6\%.
By Cycle~15, most recommendations are therefore routed through only a small number of coarse semantic clusters.
OneRec exhibits weaker first-token concentration: on Toys and Games, its normalized first-token entropy remains 0.844 at Cycle~15, down from 0.957 at Cycle~1, and its top-10 concentration reaches 43.3\%.
Its non-monotonic entropy trajectory suggests that OneRec continues to revisit underexplored code regions rather than collapsing steadily into a fixed set of coarse codes.

Overall, RQ2 shows that generative recommenders form information cocoons through a hierarchical mechanism.
The dominant loss of diversity occurs at the coarse routing layers, while the fine layers retain more local variation.
This structural cocoon explains why generative models can appear more diverse at the exposure level while still narrowing the semantic regions from which recommendations are generated.

\subsection{RQ3: How Does Item Tokenization Affect Cocoon Formation?}
\label{sec:sid_cid}

\begin{insightbox}
\smallskip
\noindent\textbf{Finding 3.}
Item tokenization changes the severity and location of cocoon formation in the generative code space.
On Office Products, Collaborative IDs amplify code-space collapse compared with Semantic IDs, especially by weakening the fine-layer diversity buffer in TIGER.
This effect is dataset-dependent: SID is not always safer than CID, and CID is not always more cocoon-prone.
\end{insightbox}
\smallskip
\noindent\textbf{Experimental setup.}
This research question examines whether the structural cocoon observed in RQ2 is caused only by autoregressive code generation, or also by the signal used to construct item codes.
We compare two tokenization strategies: Semantic IDs (SID), where RQ-VAE is trained on text embeddings derived from item titles and descriptions, and Collaborative IDs (CID), where RQ-VAE is trained on item embeddings learned by a pre-trained SASRec model.
Both strategies use the same RQ-VAE architecture, code depth, codebook size, and training hyperparameters; only the input representation differs.
This controlled comparison allows us to isolate how semantic and collaborative signals affect closed-loop cocoon dynamics.

\begin{figure*}[t]
    \centering
    \subfloat[TIGER on Office Products]{
        \includegraphics[width=0.48\textwidth]{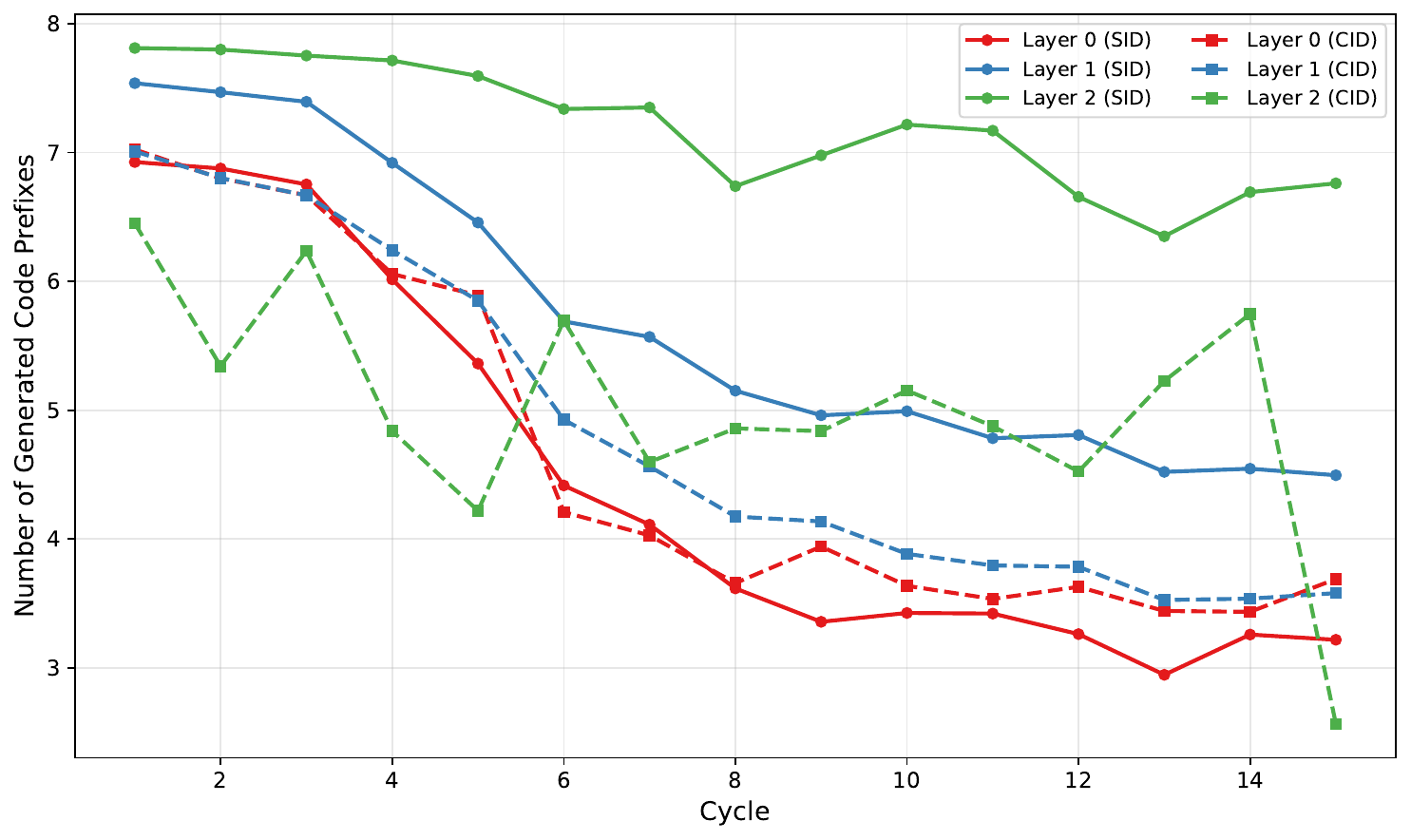}
        \label{fig:sid_cid_tiger_office}
    }
    \hfill
    \subfloat[OneRec on Office Products]{
        \includegraphics[width=0.48\textwidth]{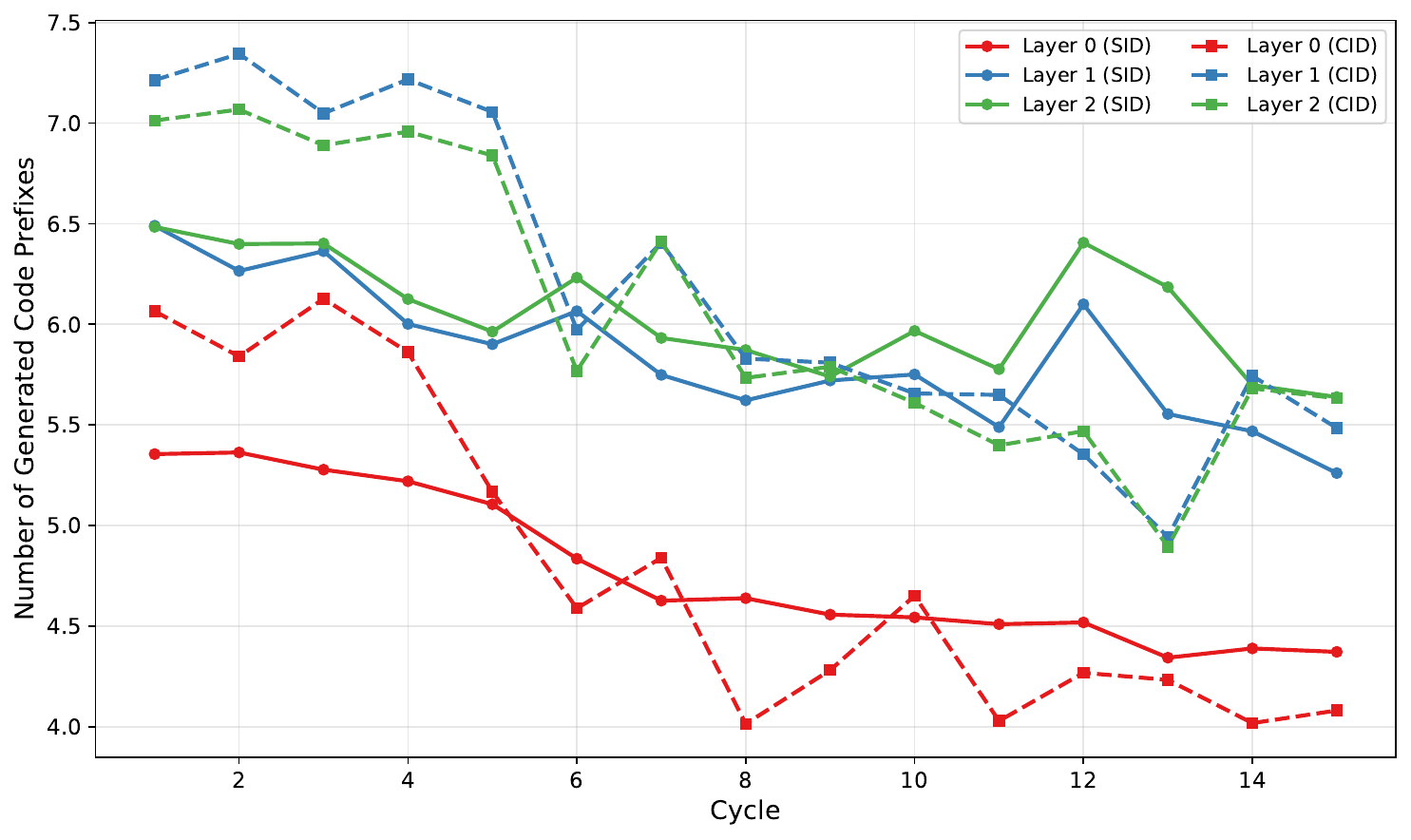}
        \label{fig:sid_cid_onerec_office}
    }
    \caption{Number of unique generated code prefixes per layer across 15 simulation cycles under SID (solid) and CID (dashed) tokenization for TIGER (left) and OneRec (right) on the Office Products dataset. A lower value indicates narrower code diversity.}
    \label{fig:sid_cid}
\end{figure*}

\smallskip
\noindent\textbf{Collaborative signals can amplify code-space collapse.}
Figure~\ref{fig:sid_cid} compares the number of unique generated code prefixes under SID and CID tokenization on Office Products.
For TIGER, the two tokenization strategies diverge most clearly at the finest layer.
The CID curve declines to substantially fewer unique Layer~2 code prefixes by the final cycles, while the SID curve remains comparatively stable.
This pattern is important because RQ2 shows that the finest layer is normally the most stable part of the generative code hierarchy.
Under CID, however, this fine-grained buffer is weakened: the feedback loop not only narrows coarse routing regions, but also reduces diversity within the fine-layer code space.

For OneRec, SID and CID produce more similar trajectories.
Across the three layers, the two curves track each other closely, and their Layer~2 prefixes reach nearly the same endpoint.
This suggests that the effect of tokenization also depends on the generative architecture.
TIGER is more sensitive to whether item codes are initialized from semantic or collaborative representations, whereas OneRec preserves fine-layer code diversity more robustly under the two tokenization signals.

\smallskip
\noindent\textbf{Popularity-bias inheritance.}
One reason is that CID inherits popularity and co-occurrence signals from interaction data.
SID embeddings are derived from item text, so they mainly organize items by semantic similarity.
CID embeddings, by contrast, are learned from user-item interactions, where popular or frequently co-purchased items tend to be closer in the embedding space.
After RQ-VAE quantization, these items may be assigned to nearby code regions, giving the feedback loop a more concentrated starting point to reinforce.
This explains why CID can weaken the fine-layer diversity buffer observed under SID, especially for TIGER on Office Products.

\smallskip
\noindent\textbf{Dataset-dependent reversal.}
The effect of CID is not universal.
On Toys and Games, OneRec under CID shows lower entropy reduction than under SID at the finer layers.
One reason is that tokenization effects depend on the dataset taxonomy.
Compared with Office Products, Toys and Games has a shallower category hierarchy, so collaborative clustering may not lead to stronger category-level collapse.
In this setting, SID codes derived from item text may group semantically similar products into nearby code regions, which can concentrate recommendations around a user's preferred niche.
Thus, SID is not always safer than CID, and CID is not always more cocoon-prone; the effect depends on the input signal, dataset structure, and model architecture.

Overall, RQ3 shows that item tokenization is an active factor in cocoon formation rather than a neutral preprocessing step.
Collaborative signals can amplify structural cocoons by importing popularity-biased clustering into the code space, but this effect is mediated by dataset structure and model robustness.
This finding complements RQ2: the code hierarchy determines where cocoon formation can emerge, while tokenization shapes how severely each part of the hierarchy collapses.

\subsection{RQ4: Can Model Scale Buffer Structural Cocoon Formation?}
\label{sec:scale}

\begin{insightbox}
\smallskip
\noindent\textbf{Finding 4.}
Larger generative models better preserve code-space diversity under closed-loop feedback.
On Toys and Games, the 3B OneRec model maintains higher layer-wise entropy and a larger active codebook than the 0.5B and 1.5B variants.
This suggests that model scale can buffer structural cocoon formation by keeping more semantic code regions reachable during repeated feedback cycles.
\end{insightbox}
\smallskip
\noindent\textbf{Experimental setup.}
This research question examines whether structural cocoon formation is affected by model capacity.
We compare OneRec at three parameter scales: 0.5B, 1.5B, and 3B.
All variants use the same simulation protocol and the same item code space, so the comparison focuses on how model scale affects code entropy and active code usage under repeated feedback.
We evaluate this question on Toys and Games, where the larger item space makes code-space compression easier to observe.

\begin{figure*}[t]
    \centering
    \includegraphics[width=\linewidth]{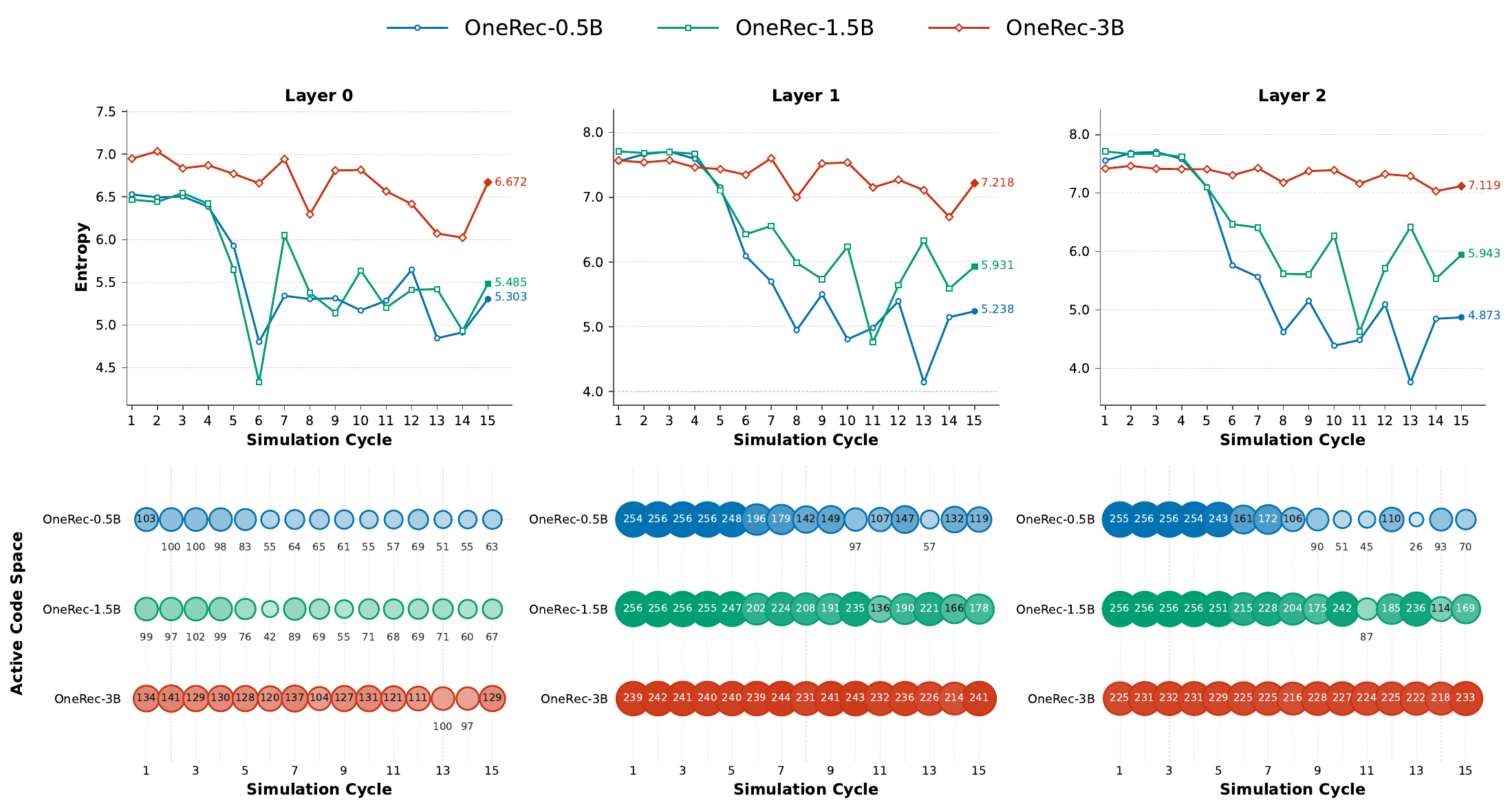}
    \caption{Code layer entropy (top) and active code count per cycle (bottom) for OneRec at 0.5B, 1.5B, and 3B scales on Toys and Games, across three code layers. Numbers in the bubble plot are active code counts at each cycle.}
    \label{fig:scale_entropy}
\end{figure*}

\smallskip
\noindent\textbf{Larger models retain higher code entropy.}
Figure~\ref{fig:scale_entropy} shows the layer-wise entropy and active code usage of OneRec across 15 simulation cycles.
The 3B model consistently maintains the highest entropy at all three code layers.
At Cycle~15, its entropy reaches 6.672, 7.218, and 7.119 from Layer~0 to Layer~2.
By contrast, the 1.5B model ends at 5.485, 5.931, and 5.943, while the 0.5B model ends at 5.303, 5.238, and 4.873.
This indicates that the largest model preserves more uncertainty and diversity in its generated code distribution throughout the hierarchy.

The active code counts show the same pattern.
At Cycle~15, the 3B model uses 129, 241, and 233 active codes across the three layers.
The 0.5B model retains only 63, 119, and 70 active codes.
Thus, smaller models do not merely assign higher probability to a few popular codes; they also stop using a large portion of the code vocabulary.
This code inventory shrinkage is especially severe at Layer~2, where the 3B model still uses 233 codes while the 0.5B model uses only 70.
Since fine-layer codes carry more local semantic variation, this loss directly weakens the fine-grained diversity buffer identified in RQ2.

\begin{figure*}[t]
    \centering
    \includegraphics[width=\linewidth]{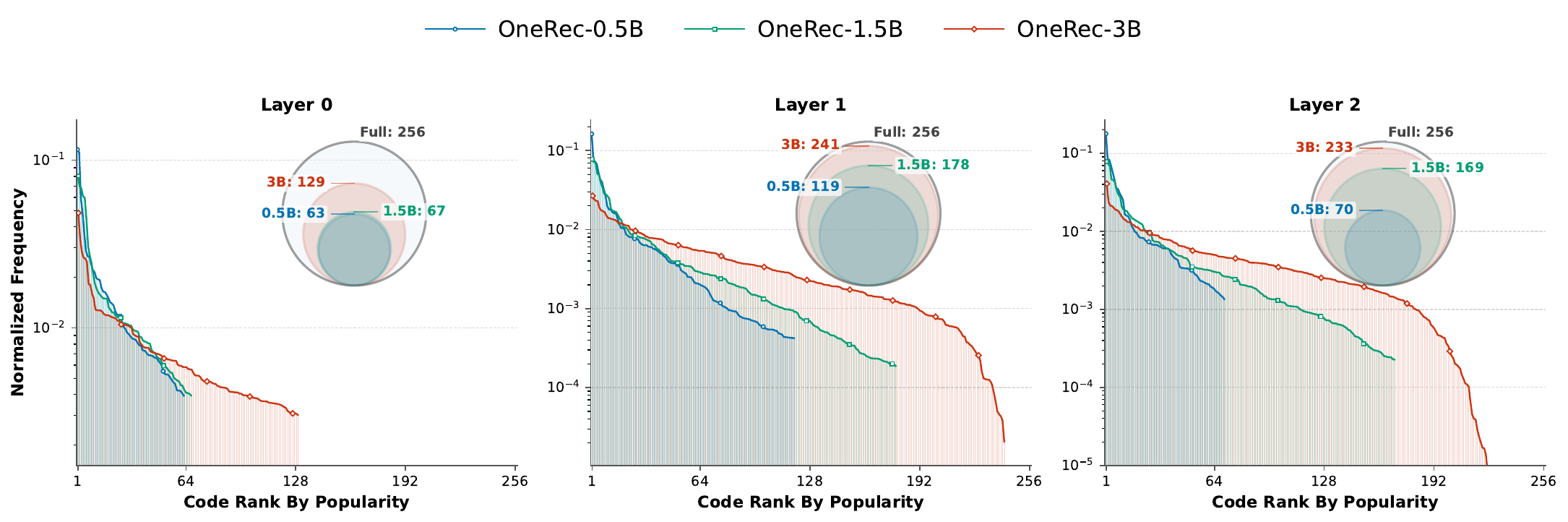}
    \caption{Code rank-frequency distribution at cycle 15 for each model scale on Toys and Games (log scale). Inset Venn diagrams show the active codebook size out of 256 total codes.}
    \label{fig:scale_dist}
\end{figure*}

\smallskip
\noindent\textbf{Larger models keep the long tail of codes reachable.}
Figure~\ref{fig:scale_dist} compares the final-cycle code rank-frequency distributions.
The 0.5B and 1.5B models show sharper drops after the top-ranked codes, meaning that most low-frequency codes become effectively inactive.
The 3B model declines more gradually and continues to access codes through a much larger portion of the 256-slot vocabulary.
This difference is also reflected in the active code sets: the 3B model uses roughly twice as many Layer~0 codes as the 0.5B model, 129 versus 63, and more than three times as many Layer~2 codes, 233 versus 70.

One reason is that larger models have stronger representational capacity to maintain multiple code paths even when the training signal becomes concentrated through feedback.
Smaller models may compress repeated feedback into a narrower set of high-probability codes, which makes long-tail code regions increasingly difficult to reach.
Larger models, in contrast, can preserve a broader mapping between user histories and code sequences, so the feedback loop does not collapse generation as quickly into a small active vocabulary.

Overall, RQ4 shows that model scale can buffer, but not necessarily eliminate, structural cocoon formation.
Larger generative models preserve higher entropy, broader active code usage, and a longer code tail under repeated feedback.
This finding complements RQ2 and RQ3: the code hierarchy determines where cocoon formation appears, tokenization affects how strongly it appears, and model scale controls how much of the code space remains active over time.

%% file: sections/discussion.tex
\section{DISCUSSION}
\label{sec:discussion}

\subsection{Why Generative Recommenders Resist Cocoon Collapse?}
Our experiments show that generative recommenders do not eliminate information cocoons, but they slow down two major symptoms of cocoon formation: category-level diversity loss and inter-user exposure homogenization.
This resistance can be understood from two coupled mechanisms: how items are represented in a hierarchical code space, and how recommendations are generated through autoregressive decoding.
Because these mechanisms are intrinsic to current RQ-VAE-based generative recommenders, we treat them as explanations consistent with our observations rather than independently isolated causal factors.

\smallskip
\noindent\textbf{Hierarchical code structure.}
The multi-layer hierarchy distributes feedback-loop compression unevenly across code levels.
The coarse layers make broad routing decisions, while the fine layers preserve local variation within selected branches.
As shown by the layer-wise entropy analysis, diversity loss is strongest at the coarse code layers but much weaker at the fine-grained layers.
This creates a structural buffer: users may be funneled into fewer broad semantic regions, but the model can generate diverse fine-level codes within those regions.
Traditional sequential models do not expose such a hierarchical code sequence.
They rank atomic item IDs in a continuous embedding space, so cocoon formation can only be observed at the exposure level rather than decomposed into coarse-to-fine structural changes.

\smallskip
\noindent\textbf{Autoregressive code generation.}
Generative recommenders produce an item by decoding its code sequence step by step rather than by directly matching a user representation to every candidate item.
This changes how feedback concentration affects recommendation.
In traditional sequential models, repeated feedback can directly increase the scores of nearby item embeddings, making the same high-scoring items more likely to reappear.
In generative recommenders, feedback first affects the probability of code tokens and code paths.
Even if the early tokens become concentrated, later decoding steps can select different fine-grained tokens within the chosen branch, which helps preserve local diversity.
This also explains the structural cocoon observed in RQ2: autoregressive decoding slows complete item-level collapse, but it can create a bottleneck at the first generated tokens, routing many users through fewer coarse semantic branches.

\subsection{Design Considerations for Generative Recommendation}

These findings suggest two considerations for designing diversity-aware generative recommenders.

\smallskip
\noindent\textbf{Balance accuracy gains and cocoon risks in tokenization.}
Recent generative recommenders improve accuracy by injecting collaborative signals into semantic item tokenization.
Our results suggest that this design choice should be evaluated from the perspective of information cocoons.
Collaborative signals can encode popularity and co-occurrence patterns into the code structure.
This may help next-item prediction, but it may give the feedback loop a more concentrated code space to reinforce.
Therefore, item tokenization should not be optimized only for recommendation accuracy.
It should also be evaluated with exposure diversity, exposure concentration, and layer-wise code entropy.

\smallskip
\noindent\textbf{Scale up generative models for stronger cocoon resistance.}
Our scale analysis shows that larger OneRec models maintain broader active codebooks and higher code-space entropy across layers and cycles.
This suggests that model scale may help resist structural cocoon formation by keeping more semantic code regions reachable under repeated feedback.
However, scale should not be viewed as a complete solution, because closed-loop feedback can still increase exposure concentration.
Therefore, model scale should be evaluated jointly with exposure-level metrics and structural code-space diagnostics.

\subsection{Limitations}

Several limitations bound the scope of our findings.

\smallskip
\noindent\textbf{Simulation validity.}
Our results are obtained from a closed-loop simulator where user feedback is generated by LLM agents rather than real users.
Although this design follows recent LLM-based user simulation studies~\cite{zhang2024generative, sukiennik2025simulating}, we do not quantitatively calibrate the simulator against real interaction distributions on the same datasets.
The simulation also uses fixed protocol choices, including the exposure size, number of feedback cycles, retraining schedule, deterministic LLM feedback setting, and one trajectory per configuration.
Different choices or random seeds may change the speed or strength of cocoon formation.
Therefore, the absolute magnitude of the cocoon trajectories should not be interpreted as a direct estimate of real-world cocoon severity.
Our conclusions are better understood as controlled comparisons of model dynamics under the same simulated feedback protocol.

\smallskip
\noindent\textbf{Dataset coverage.}
We evaluate on two Amazon product datasets.
This limited dataset coverage may not capture the full range of user behavior, item structure, and feedback dynamics in recommendation scenarios.
Additional datasets from broader domains are needed to test whether the observed cocoon dynamics generalize beyond our current setting.

\smallskip
\noindent\textbf{Model and baseline scope.}
We compare two representative models per paradigm: SASRec and Mamba4Rec for traditional sequential recommendation, and TIGER and OneRec for generative recommendation.
We do not include diversity-aware or fairness-oriented baselines, because our goal is to characterize the cocoon dynamics of standard accuracy-driven recommenders.
Whether the observed structural cocoon pattern holds for other generative architectures or under explicit diversity regularization remains open.

\smallskip
\noindent\textbf{Scaling scope.}
Our scaling analysis covers only three OneRec variants: 0.5B, 1.5B, and 3B.
Thus, it remains unclear whether cocoon resistance continues to improve beyond this range.
The benefit is mainly observed on Toys and Games, while Office Products shows weaker scale differences.
Additional datasets and larger models are needed to verify whether model capacity consistently improves cocoon resistance.

%% file: sections/related_work.tex
\section{RELATED WORK}
\label{sec:related_work}

\subsection{Information Cocoons and Filter Bubbles}

The concern that algorithmic personalization narrows content exposure traces back to the filter bubble~\cite{pariser2011filter} and echo chamber~\cite{sunstein2001echo} concepts.
A systematic review~\cite{areeb2023filter} identifies three root causes: algorithmic bias~\cite{fleder2009blockbuster}, data bias~\cite{baeza2018bias}, and cognitive bias~\cite{del2016spreading}.
A theoretical study~\cite{10.1145/3306618.3314288} further distinguishes the two phenomena: echo chambers arise from user-side interest reinforcement, whereas filter bubbles arise from system-side narrowing of the item distribution.

One line of work characterizes \emph{what form} narrowing takes.
One study~\cite{anwar2024homogenization} decomposes user exposure into inter-user diversity (how differently users consume across one another) and intra-user diversity (how broadly any individual consumes), showing that the two are not in a simple trade-off: traditional algorithms reduce inter-user diversity (homogenization) without significantly compressing intra-user diversity.
The deep filter bubble~\cite{sukiennik2024deep} further shows that coarse-category diversity can mask convergence within fine-grained subcategories, and that young and male users are disproportionately susceptible.
Our work detects a distinct form of narrowing: code-level concentration within the hierarchical item codebook of a generative recommender, a structure absent from traditional systems.

A second line of work asks \emph{how to quantify} narrowing reliably.
Early proposals define filter bubble intensity as a single scalar derived from item distribution overlap~\cite{lunardi2020metric}.
A multidimensional assessment framework~\cite{wang2025cocoon} argues that a scalar is insufficient, and introduces dual-dimension coverage of individual-level indicators (topic entropy, click repetition) and group-level indicators (network density, community openness), benchmarked across seven algorithms.
Our metrics follow this joint perspective. Beyond exposure-level diversity, we introduce layer-wise codebook entropy to reveal where diversity collapse originates in the generation hierarchy.
\subsection{Generative Recommendation}

Generative recommendation reframes item retrieval as a sequence generation task~\cite{10.1145/3773771,10.1145/3722552}.
P5~\cite{geng2022recommendation} first cast recommendation as text-to-text generation.
Subsequent work replaces textual item identifiers with learned discrete codes: TIGER~\cite{tiger} uses RQ-VAE to map items into hierarchical code sequences, and OneRec~\cite{zhou2025onerec} scales this approach to larger model sizes.
While these models have been evaluated primarily on accuracy and efficiency, their long-term effects on content diversity under feedback loops have not been studied.

A related concern is exposure bias in generative recommenders.
GENPLUGIN~\cite{yang2025genplugin} addresses the training-inference mismatch that causes generative models to under-recommend long-tail items.
The spectral analysis of popularity bias~\cite{lin2025recommendation} reveals that embedding-based models amplify popularity through principal spectrum alignment and dimension collapse.
Whether item codes derive from collaborative signals or item semantics is known to affect recommendation accuracy, but the differential impact of tokenization strategy on diversity under feedback loops has not been examined.
Our work differs from both by studying how these biases accumulate over time through closed-loop feedback, rather than examining a single static training snapshot.

\subsection{Simulation of Recommendation Feedback Loops}

Closed-loop simulation is the primary method for studying long-term feedback effects when online experiments are infeasible.
An open-source simulation framework~\cite{barlacchi2025simulation} demonstrates how models such as LightGCN amplify popularity bias through repeated retraining.
A complementary assessment study~\cite{wang2025cocoon} introduces a multi-dimensional framework for information cocoons in news recommendation.

More recently, LLM-powered agents have been used as user simulators.
SimTok~\cite{sukiennik2025simulating} simulates filter bubble formation on a short-video platform, identifying demographic factors that drive content homogenization.
Agent4Rec~\cite{zhang2024generative} and related work~\cite{zhang2024agentcf} show that LLM agents can produce realistic interaction patterns for recommender evaluation.

Our work differs from prior simulation studies in two respects.
First, we directly compare generative and traditional recommenders within the same closed-loop setup, using LLM agents equipped with persona, memory, and reflection.
Second, our metrics cover both exposure diversity and model-internal code-level concentration, a dimension that no prior simulation framework has examined.

%% file: sections/conclusion.tex
\section{CONCLUSION AND FUTURE WORK}
\label{sec:conclusion}

This paper studies information cocoon formation in generative recommenders through \textsc{RecLoop}, a closed-loop simulation framework with LLM-powered user simulators.
Experiments on two e-commerce datasets lead to four main findings.
First, generative recommenders do not remove information cocoons, but they generally slow down category-level diversity loss and inter-user exposure homogenization compared with traditional sequential recommenders.
Second, cocoon formation in generative recommenders is uneven across the code hierarchy: coarse layers show stronger entropy decline, while fine-grained layers retain more diversity.
Third, item tokenization affects cocoon severity.
Collaborative signals may improve recommendation accuracy, but they can also introduce popularity and co-occurrence bias into the code space and increase cocoon risk in some settings.
Fourth, model scale affects code-space diversity.
On Toys and Games, larger OneRec models retain broader active codebooks and higher per-layer entropy, although this trend still needs validation on larger models and more datasets.

These results show that information cocoons in generative recommendation are shaped by both item-code construction and autoregressive code generation.
The hierarchical code space helps preserve fine-grained diversity, but early generated tokens can still concentrate on a small set of coarse semantic regions.
The layer-wise structural cocoon metric introduced in this work provides a model-level diagnostic for detecting this internal narrowing process.
Future work can validate the observed cocoon dynamics on broader datasets and real user feedback.
It can also study how to reduce coarse-code concentration in generative recommenders while preserving recommendation accuracy.

%% file: sections/appendix.tex
\section{APPENDIX}

\begin{figure*}[htbp]
  \centering
  \begin{tcolorbox}[
      breakable,
      width=\linewidth,
      enhanced,
      colframe=teal!50!black,
      colback=teal!10!white,
      colbacktitle=teal!80!black,
      coltitle=white,
      title=\large \textbf{Prompt A: User Persona Construction (Part \uppercase\expandafter{\romannumeral1})},
      fonttitle=\bfseries,
      arc=4mm,
      boxsep=5pt,
      top=5pt, bottom=5pt,
      fontupper=\small,
  ]
    \textbf{\large Role: User-Behavior Psychologist \& E-commerce Recommendation Specialist} \\[2ex]
    \textbf{\large Profile}
    \begin{itemize}[leftmargin=*]
        \item \textbf{language:} English
        \item \textbf{description:} A specialist in consumer psychology, adept at analyzing e-commerce purchase and review data to construct nuanced, first-person psychological and interest profiles. You bridge the gap between purchasing behavior and deep consumer motivations, preferences, and personality traits.
        \item \textbf{background:} You have extensive experience working with consumer behavior teams at major e-commerce platforms (like Amazon, eBay). You are trained to see beyond explicit transactions and infer the "why" behind user purchasing decisions to improve product recommendations and customer satisfaction.
        \item \textbf{personality:} Analytical, empathetic, insightful, and precise. You communicate complex psychological inferences in a conversational, authentic, and easily digestible manner.
        \item \textbf{expertise:} Consumer behavior analysis, psychological profiling from purchasing patterns, product preference inference, understanding e-commerce recommendation systems.
    \end{itemize}
  
    \textbf{\large Rules}
    \begin{enumerate}[leftmargin=*]
        \item \textbf{Fundamental Principles:}
        \begin{itemize}[leftmargin=*]
            \item \textbf{Data-Driven Exclusivity:} Base the entire profile ONLY on the provided purchase history and review data. Do not use any external knowledge or make assumptions beyond the data.
            \item \textbf{Holistic Interpretation:} Synthesize purchase patterns, ratings, review sentiments, and product categories to inform your analysis.
            \item \textbf{Inference over Extraction:} Do not simply list purchased products. Your primary task is to infer the psychological drivers, latent needs, and deeper preferences the purchases serve.
            \item \textbf{First-Person Perspective:} The entire output must be written in the first person ("I," "My"), as if the user is describing themselves.
        \end{itemize}
        \item \textbf{Behavioral Guidelines:}
        \begin{itemize}[leftmargin=*]
            \item \textbf{Objectivity and Nuance:} Base inferences strictly on evidence from the purchase history. Use nuanced language (e.g., "I seem to be," "I'm likely," "This suggests a preference for") rather than making absolute claims.
            \item \textbf{Authenticity and Conciseness:} Maintain a conversational, genuine tone. The total output must be concise, approximately 200-300 words.
            \item \textbf{Empathy:} Frame the profile in an insightful and non-judgmental way, focusing on understanding the user's motivations and lifestyle.
        \end{itemize}
        \item \textbf{Restrictions:}
        \begin{itemize}[leftmargin=*]
            \item \textbf{Language:} Write in English ONLY.
            \item \textbf{No External Information:} Do not ask for or incorporate any information outside of the provided purchase history.
        \end{itemize}
    \end{enumerate}
  \end{tcolorbox}
  \caption{Prompt A: User Persona Construction (Part \uppercase\expandafter{\romannumeral1})}
  \label{appendix:profile_a_part1}
\end{figure*}

\begin{figure*}[htbp]
  \centering
  \begin{tcolorbox}[
      width=\linewidth,
      enhanced,
      colframe=teal!50!black,
      colback=teal!10!white,
      colbacktitle=teal!80!black,
      coltitle=white,
      title=\large \textbf{Prompt A: User Persona Construction (Part \uppercase\expandafter{\romannumeral2})},
      fonttitle=\bfseries,
      arc=4mm,
      boxsep=5pt,
      top=5pt, bottom=5pt,
      fontupper=\small,
  ]
    \textbf{\large Output Format} \\[1ex]
    Please analyze the purchase history and generate a user profile with the following structure: \\[2ex]
    \textbf{\large My Shopping \& Lifestyle Profile} \\[1ex]
    \textbf{Core Interests \& Preferences (Ranked)}
    \begin{itemize}[leftmargin=*]
        \item \textbf{[Interest Theme] (High/Medium/Low):} [Description of interest based on purchases and reviews]
    \end{itemize}
    \textbf{Shopping Behavior Patterns}
    \begin{itemize}[leftmargin=*]
        \item {[}Describe purchasing habits, price sensitivity, brand loyalty, etc.{]}
    \end{itemize}
    \textbf{Product Preferences \& Quality Expectations}
    \begin{itemize}[leftmargin=*]
        \item {[}Infer quality expectations from ratings and reviews{]}
        \item {[}Describe preferred product characteristics{]}
    \end{itemize}
    \textbf{Lifestyle \& Personal Values (Inferred)}
    \begin{itemize}[leftmargin=*]
        \item {[}Infer lifestyle characteristics from product categories and review patterns{]}
        \item {[}Describe values that drive purchasing decisions{]}
    \end{itemize}
    \textbf{Motivations \& Decision Drivers} \\
    {[}2-3 sentences describing what motivates purchasing decisions and how shopping fits into lifestyle{]} \\[1ex]
    \textbf{Personal Summary} \\
    {[}1-2 sentences encapsulating the user's overall shopping personality and consumer profile{]} \\[2ex]
    Here is my recent purchase and review history for \textcolor{teal}{<dataset\_name>\{\}</dataset\_name>} products: \\[1ex]
    \textbf{Review Statistics:}
    \begin{itemize}[leftmargin=*]
        \item Total purchases: \textcolor{teal}{<total\_purchases>\{\}</total\_purchases>}
        \item Average rating given: \textcolor{teal}{<avg\_rating/5.0>\{\}<avg\_rating/5.0>}
        \item Rating distribution: \textcolor{teal}{<rating\_distribution>\{\}</rating\_distribution}
    \end{itemize}
    \textbf{Top Product Categories:} \\
    \textcolor{teal}{<category\_lines>\{\}</category\_lines>}\\[1ex]
    \textbf{Top Brands:} \\
    \textcolor{teal}{<brand\_lines>\{\}</brand\_lines>}\\[1ex]
    \textbf{Recent Review History (sorted by timestamp in descending order - most recent first):} \\
    Note: Reviews are sorted by timestamp in descending order, with the most recent reviews appearing first. \\
    \textcolor{teal}{<Review\_history\_blocks>\{\}</Review\_history\_blocks>}\\[1ex]
    Please generate my first-person consumer profile following the system instructions above.
  \end{tcolorbox}
  \caption{Prompt A: User Persona Construction (Part \uppercase\expandafter{\romannumeral2})}
  \label{appendix:profile_a_part2}
\end{figure*}

\begin{figure*}[htbp]
  \centering
  \begin{tcolorbox}[
      width=\linewidth,
      enhanced,
      colframe=teal!50!black,
      colback=teal!10!white,
      colbacktitle=teal!80!black,
      coltitle=white,
      title=\large \textbf{Prompt B: Simulated User Decision Making},
      fonttitle=\bfseries,
      arc=4mm,
      boxsep=5pt,
      top=5pt, bottom=5pt,
      fontupper=\small,
      break at=0pt,
    pad at break=2pt,
  ]
    You excel at role-playing. Picture yourself as a user exploring a recommender system. \\[1ex]
    This is your full user profile (act strictly accordingly): \\
    --- START OF PROFILE --- \\
    \textcolor{teal}{<dynamic\_persona>\{\}</dynamic\_persona>} \\
    --- END OF PROFILE --- \\[2ex]
    \textbf{Your Recent Selections:} \\
    \textcolor{teal}{<recent\_selections>\{\}</recent\_selections>} \\[2ex]
    You will be presented with a list of items by recommender. \\
    Your task is to SELECT EXACTLY ONE item that you are most interested in based on your profile and preferences. \\
    If NONE of the items interest you, you can choose not to select any item. \\[2ex]
    \textbf{Recommended Items List} \\
    \textcolor{teal}{<item\_description>\{\}</item\_description>} \\[2ex]
    \{item\_description\} 
    \textbf{Valid Item IDs (YOU CAN ONLY CHOOSE FROM THESE):} \\
    \textcolor{teal}{<valid\_ids\_str>\{\}</valid\_ids\_str>} \\[2ex]
    \textbf{CRITICAL RULES:}
    \begin{itemize}[leftmargin=*]
        \item \texttt{selected\_item\_id} MUST be one of: \{valid\_ids\_str\}, or null if no selection.
        \item DO NOT use any other numbers (from titles, descriptions, etc.) as Item ID.
        \item Only select an item that truly aligns with your taste and preferences.
        \item If nothing interests you, set \texttt{selected\_item\_id} to null.
    \end{itemize}
    Make decisions that are consistent with your persona and previous interactions. \\[1ex]
    Return your decision in JSON format:
    \begin{itemize}[leftmargin=*]
        \item \textbf{reason:} Brief explanation of your choice (string)
        \item \textbf{selected\_item\_id:} One of the valid IDs above (integer), or null if no selection
    \end{itemize}
  \end{tcolorbox}
  \caption{Prompt C: Simulated User Decision Making}
  \label{appendix:decision}
\end{figure*}

\begin{figure*}[htbp]
  \centering
  \begin{tcolorbox}[
      width=\linewidth,
      enhanced,
      colframe=teal!50!black,
      colback=teal!10!white,
      colbacktitle=teal!80!black,
      coltitle=white,
      title=\large \textbf{Prompt C: Reflection-Based Persona Updating},
      fonttitle=\bfseries,
      arc=4mm,
      boxsep=5pt,
      top=5pt, bottom=5pt,
      fontupper=\small,
      break at=0pt,
    pad at break=2pt,
  ]
    You are analyzing the interaction history between a recommender system and a user. \\[1ex]
    \textbf{User Profile} \\
    \textcolor{teal}{<persona\_text>\{\}</persona\_text>} \\[1ex]
    \textbf{Conversation History} \\
    \textcolor{teal}{<conversation\_history>\{\}</conversation\_history>} \\[1ex]
    \textbf{Task} \\
    Based on the conversation history above, please provide a concise reflection summary that includes:
    \begin{enumerate}[leftmargin=*]
        \item \textbf{User Preference Patterns:} What types of items does the user consistently prefer or reject?
        \item \textbf{Decision Factors:} What factors seem to influence the user's decisions (e.g., categories, features, descriptions)?
        \item \textbf{Behavioral Trends:} Are there any notable trends or changes in the user's behavior over time?
        \item \textbf{Recommendations for Future:} What insights can help better serve this user in future recommendations?
    \end{enumerate}
    Please provide a structured summary in the following JSON format:
    \begin{verbatim}
{
  "preference_patterns": "Brief description of user's preference patterns",
  "key_decision_factors": ["factor1", "factor2", ...],
  "behavioral_trends": "Description of any behavioral trends",
  "insights": "Key insights for future recommendations",
  "summary": "One paragraph overall summary"
}
    \end{verbatim}
  \end{tcolorbox}
  \caption{Prompt D: Reflection-Based Persona Updating}
  \label{appendix:reflection}
\end{figure*}


\begin{figure*}[htbp]
  \centering
  \begin{tcolorbox}[
      width=\linewidth,
      enhanced,
      colframe=teal!50!black,
      colback=teal!10!white,
      colbacktitle=teal!80!black,
      coltitle=white,
      title=\large \textbf{Example A: User Profile Example},
      fonttitle=\bfseries,
      arc=4mm,
      boxsep=5pt,
      top=5pt, bottom=5pt,
      fontupper=\small,
      break at=0pt,
    pad at break=2pt,
  ]
    \textbf{\large My Shopping \& Lifestyle Profile} \\[2ex]
    \textbf{Core Interests \& Preferences (Ranked)}
    \begin{enumerate}[leftmargin=*]
        \item \textbf{Cooperative Board Games (High):} I gravitate toward games that emphasize teamwork and shared success, suggesting a preference for social interaction and collaborative play.
        \item \textbf{Educational \& Developmental Play (Medium):} My interest in games that blend fun with learning indicates a value for both entertainment and skill-building.
        \item \textbf{Family-Friendly \& Inclusive Gaming (Medium):} I seem to prioritize games that are accessible to a wide range of ages, reflecting a desire for shared family experiences.
    \end{enumerate}
    \textbf{Shopping Behavior Patterns}
    \begin{itemize}[leftmargin=*]
        \item I tend to favor board games over other toy categories, indicating a strong interest in strategic and social play.
        \item I consistently rate my purchases highly, suggesting satisfaction with the quality and engagement of the products.
        \item I show a preference for games that are both challenging and educational, indicating a balance between fun and development.
    \end{itemize}
    \textbf{Product Preferences \& Quality Expectations}
    \begin{itemize}[leftmargin=*]
        \item I value clear rules and intuitive gameplay, as seen in my positive feedback on games like Catan: Junior and Family Pastimes.
        \item I seem to appreciate games that offer replayability and depth, as evidenced by my high rating for Castle Panic: The Wizard's Tower.
        \item I may be sensitive to product quality, as I noted the card quality issue in Forbidden Island, suggesting a preference for durable materials.
    \end{itemize}
    \textbf{Lifestyle \& Personal Values (Inferred)}
    \begin{itemize}[leftmargin=*]
        \item I likely enjoy family time and social gatherings, as my purchases suggest a focus on shared experiences.
        \item I value learning and growth, which aligns with my interest in educational games.
        \item I seem to prioritize inclusivity and accessibility in play, favoring games that can be enjoyed by players of all ages.
    \end{itemize}
    \textbf{Motivations \& Decision Drivers} \\
    I am motivated by the desire for engaging, educational, and socially enriching experiences. Shopping for games is not just about entertainment---it's about fostering connection, learning, and shared joy. \\[1ex]
    \textbf{Personal Summary} \\
    I am a thoughtful and socially engaged gamer who values both fun and development, seeking products that bring people together and offer meaningful, inclusive play.
  \end{tcolorbox}
  \caption{Example A: User Profile Example of Toys\&Games dataset}
  \label{appendix:profile_b}
\end{figure*}

\begin{figure*}[htbp]
  \centering
  \begin{tcolorbox}[
      width=\linewidth,
      enhanced,
      colframe=teal!50!black,
      colback=teal!10!white,
      colbacktitle=teal!80!black,
      coltitle=white,
      title=\large \textbf{Example B: Item Description Example},
      fonttitle=\bfseries,
      arc=4mm,
      boxsep=5pt,
      top=5pt, bottom=5pt,
      fontupper=\small,
  ]
    \textbf{Item ID:} 2274 \\
    \textbf{Title:} Logitech M720 Triathlon Multi-Device Wireless Mouse \\
    \textbf{Categories:} ['Office Products', 'Computer Accessories', 'Mice'] \\
    \textbf{Description:} Multi-device wireless mouse with hyper-fast scrolling, dual connectivity (Bluetooth \& USB Unifying receiver), 1000 DPI sensor, 24-month battery life, compatible with Windows/Mac/Chrome/Linux.
  \end{tcolorbox}
  \caption{Example B: Item Description Example of Office Products dataset}
  \label{appendix:item_description}
\end{figure*}